\documentclass[twocolumn]{aastex62}
\pdfoutput=1
\date{\today}
\usepackage{graphicx}
\usepackage[utf8]{inputenc}
\usepackage{amsmath,amstext}
\usepackage[T1]{fontenc}
\usepackage{apjfonts} 
\usepackage[figure,figure*]{hypcap}
\usepackage{amssymb}
\usepackage{color}
\usepackage{hyperref}
\usepackage{longtable}
\newlength\figureheight
\newlength\figurewidth
\usepackage[caption=false]{subfig}

\usepackage[amssymb,mediumspace,thickqspace]{SIunits}

\shorttitle{Parsec dusty winds}
\shortauthors{Leftley J.H. et al.}

\begin{document}
\author[0000-0001-6009-1803]{James H. Leftley}
\affiliation{Department of Physics \& Astronomy, University of Southampton, Southampton, SO17 1BJ, UK}

\author[0000-0002-6353-1111]{Sebastian F. H\"onig}
\affiliation{Department of Physics \& Astronomy, University of Southampton, Southampton, SO17 1BJ, UK}

\author[0000-0003-0220-2063]{Daniel Asmus}
\affiliation{Department of Physics \& Astronomy, University of Southampton, Southampton, SO17 1BJ, UK}

\author[0000-0001-8281-5059]{Konrad R. W. Tristram}
\affiliation{European Southern Observatory, Alonso de Córdova 3107, Casilla 19001, Santiago, Chile}


\author[0000-0003-3105-2615]{Poshak Gandhi}
\affiliation{Department of Physics \& Astronomy, University of Southampton, Southampton, SO17 1BJ, UK}

\author[0000-0002-2216-3252]{Makoto Kishimoto}
\affiliation{Department of Astrophysics and Atmospheric Sciences, Kamigamo-Motoyama, Kita-ku, Kyoto, 603-8555, Japan}

\author[0000-0002-9205-4887]{Marta Venanzi}
\affiliation{Department of Physics \& Astronomy, University of Southampton, Southampton, SO17 1BJ, UK}

\author[0000-0002-5168-7979]{David J. Williamson}
\affiliation{Department of Physics \& Astronomy, University of Southampton, Southampton, SO17 1BJ, UK}

\title{Parsec-scale Dusty Winds in Active Galactic Nuclei: Evidence for Radiation Pressure Driving\footnote{Based on European Southern Observatory (ESO) observing programmes 71.B-0062, 078.B-0303, 083.B-0288, 083.B-0452, 084.B-0366, 086.B-0019, 086.B-0242, 087.B-0018, 087.B0401, 091.B-0025, 091.B-0647, 092.B-0718, 092.B-0738, 093.B-0287, 093.B-0647, 094.B-0918, 095.B-0376,0101.B-0411 and 290.B-5113 as well as the Spitzer programmes 86, 526, 3269, 3605, 30572, and 50588.}}

\begin{abstract}

Infrared interferometry of local AGN has revealed a warm ($\sim$300K-400K) polar dust structure that cannot be trivially explained by the putative dust torus of the unified model. This led to the development of the disk+wind scenario which comprises of a hot ($\sim$1000K) compact equatorial dust disk and a polar dust wind. This wind is assumed to be driven by radiation pressure and, therefore, we would expect that long term variation in radiation pressure would influence the dust distribution. In this paper we attempt to quantify if and how the dust distribution changes with radiation pressure. We analyse so far unpublished VLTI/MIDI data on 8 AGN and use previous results on 25 more to create a sample of 33 AGN. This sample comprises all AGN successfully observed with VLTI/MIDI. For each AGN, we calculate the Eddington ratio, using the intrinsic 2-10\,keV X-ray luminosity and black hole mass, and compare this to the resolved dust emission fraction as seen by MIDI. We tentatively conclude that there is more dust in the wind at higher Eddington ratios, at least in type 2 AGN where such an effect is expected to be more easily visible.
\end{abstract}


\section{Introduction}
\label{Intro}

The introduction of the MID-infrared Interferometer \citep[MIDI,][]{leinert_midi_2003} to the Very Large Telescope Interferometer (VLTI) has granted the Active Galactic Nuclei (AGN) community access to unprecedented spatial scales. MIDI operated in the N-band which in most galaxies is dominated by emission from warm, $300-400$K, dust. Direct observation of this dust was achieved on multiple AGNs, the most iconic of which were the Circinus galaxy, NGC\,1068, and NGC\,3783 \citep[e.g][]{kishimoto_mapping_2011,honig_parsec-scale_2012,honig_dust_2013,honig_dust-parallax_2014,burtscher_diversity_2013,tristram_dusty_2014,lopez-gonzaga_revealing_2014,lopez-gonzaga_mid-infrared_2016}.

The spatial scales covered by MIDI provided a direct view of the expected location of the putative warm clumpy dust torus predicted by the classical model of AGN \citep{antonucci_unified_1993}. The bulk of the dust mass in the torus thermally emits in the mid-IR and, therefore, should be visible to MIDI. It was thought that the dust torus would be seen as an equatorial extension in a type 2 Seyfert (Sy2) and either as a mild equatorial or radially symmetric extension in a type 1 Seyfert (Sy1) \citep[e.g.][]{schartmann_towards_2005,schartmann_three-dimensional_2008,stalevski_3d_2012,honig_parsec-scale_2012}.

MIDI found extended dust emission on the angular scale of the dust torus. However, contrary to expectations, the emission arose from the polar regions of the AGN instead. So far, 8 AGN, 2 Sy1s and 6 Sy2s, have been studied in enough detail to constrain non-radially symmetric geometry and none have shown the equatorial extension expected of the dust torus \citep[e.g.][]{lopez-gonzaga_mid-infrared_2016,leftley_new_2018}. The 2 Sy1s and 4 of the Sy2s show a polar extension and the remaining 2 Sy2s are radially symmetric. When modelling the interferometric data, it was found that the emission can be interpreted as a combination of two distinct sources. One source is partially resolved, and often polar extended, and the other is unresolved at all baseline lengths available. This, however, is not the only interpretation of the emission, it can also be explained as single power-law component, e.g. \citet{kishimoto_mapping_2011} and \citet{honig_parsec-scale_2012}. In this work, we use the two-component model. This allows us to separate the extended and unresolved emission and makes our work easily comparable to \citet{burtscher_diversity_2013} and \citet{lopez-gonzaga_mid-infrared_2016}.

\citet{honig_dust_2013} used a polar dust wind to explain this extension and replaced the putative dust torus with a geometrically thin hot, $\approx1000$\,K, dust disk to act as the dust reservoir and source of obscuration in the unified model of AGN \citep{antonucci_unified_1993}. This wind is thought to originate from just outside the sublimation radius where dust disk could be puffed up due to radiation pressure and launched as a dusty wind. This would form a hollow cone-like structure around the Narrow Line Region (NLR). Not only does such an extended structure connect the smallest scales around the central engines with the larger scale host galaxies, it is also capable of explaining the apparent near isotropy of the mid-infrared emission inferred in large AGN samples \citep[e.g][]{gandhi_resolving_2009, asmus_subarcsecond_2015}, all observations of the Circinus galaxy \citep{stalevski_dissecting_2017,vollmer_thick_2018,stalevski_dissecting_2019}, and the larger scale polar dust in \citet{asmus_subarcsecond_2016}.

Radiation pressure, specifically IR radiation, was originally a way to explain how the geometrically thick classical dusty torus could be supported against collapsing into a thin disk \citep{pier_radiation-pressure--supported_1992}. In modelling from \citet{chan_radiation-driven_2016,chan_geometrically_2017} it was found that, in the low Eddington ratio case of Seyfert AGN, the vertical support from the radiation is insufficient and the dust collapses into a thin disk but, nevertheless, a radial dusty wind is produced. Hydrodynamic models \citep[e.g.][]{wada_obscuring_2015,namekata_sub-parsec-scale_2016,williamson_3d_2019} came to similar conclusions although the geometry of the disk and wind varies from model to model depending on the included model parameters and resolution of the simulation. \citet{williamson_3d_2019} make the further point that the wind was highly sensitive to the anisotropy of the central radiation and to the Eddington ratio. Therefore, we should see a change in the dust distribution in AGN with Eddington ratio.

In the study of ESO\,323-G77 it was inferred that the unresolved emission, as seen by MIDI, is the hot dust close to the sublimation radius \citep{leftley_new_2018}. This had been concluded for other objects previously \citep[e.g.][]{burtscher_dust_2009}, however, ESO\,323-G77 was the first polar extended source where the unresolved component is dominant. Its dust structure, from SED modelling, is very compact; the compact emission is similar to that seen in two quasars \citep{kishimoto_mapping_2011}. Because ESO\,323-G77 is considered to have a higher Eddington ratio than typical Seyferts, the tentative conclusion was drawn that more dust resides in the disk at higher Eddington ratios.

In this paper we present a detailed analysis of 8 AGN that have so far unpublished MIDI data. We will use the larger sample of AGNs, comprised of the 8 new objects and the 25 from literature \citep{burtscher_diversity_2013,lopez-gonzaga_mid-infrared_2016,fernandez-ontiveros_embedded_2018,fernandez-ontiveros_compact_2019}, to investigate the tentative evolution of the dust distribution with Eddington ratio put forward previously in \citet{leftley_new_2018} and, separately, conclude whether any of these objects invite further study with the new era instrumentation.

In Section \ref{S:Observations} we describe the observations used in this paper and our reduction method. In Section \ref{Modelling} we detail the modelling performed on the interferometric data. Next, in Section \ref{S:Results} we present the results of each object and our investigation into the relationship between Eddington ratio and dust distribution. In Section \ref{Discussion} we discuss the implications of our findings in Section \ref{S:Results} and possible avenues of future work. Finally, in Section \ref{S:Summary} we summarise our work.

\section{Observations and Reduction}
\label{S:Observations}

We analyse all yet unpublished MIDI data on 8 objects: Fairall\,51, Fairall\,49, Fairall\,9, MCG-06-30-15, Mrk\,509, NGC\,2110, NGC\,7213, and NGC\,7674; which we supplement with archival and new spectrophotometric data obtained with the VLT-mounted Imager and Spectrometer for the mid-IR (VISIR; \citealt{lagage_successful_2004}) as described in detail in the following.

\subsection{MIDI}

We use all publicly available MIDI data on the ESO archives for each object, including previously unpublished data. Observation dates and programme numbers are listed in Table \ref{tab:files}. We reduced the observations in the same manner as \citet{leftley_new_2018}. The observations are grouped into 30 minute blocks in accordance with \citet{lopez-gonzaga_mid-infrared_2016}, reduced using \textsc{ews} from \textsc{mia+ews\,2.0} \citep{burtscher_observing_2012}, and calibrated using the closest standard star in time. We do not, however, use the structure function (SF) method for estimating the error from atmospheric variation. This is because the SF for all objects do not show useful structure. This is due, in part, to an insufficient number of calibrators observed during the night and possibly further influenced by the introduced variation from no-track calibrators (see Section \ref{Subsection: No track}). From the final product we extracted the correlated flux and compared it to the total flux from VISIR single dish spectrophotometry (see Section \ref{Section:VISIR}).

\subsubsection{No-Track Mode Calibration}
\label{Subsection: No track}
The no-track mode on MIDI, used by many of the observations in this paper, can cause the measured correlated flux to be artificially reduced if the object fringe strays too far from the set delay position. To prevent this, \textsc{ews} only selects those frames where the fringes are within 100$\,\mu$m of the set delay position. This behaviour is standard for all modes offered by MIDI. To check for any systematic differences between tracked and non tracked objects we compared the normalised difference in ADU of calibrators that have adjacent non tracked and tracked observations. Figure \ref{F: ADU comp multi calib} in the Appendix shows the objects used as well as the normalised ADU differences at 8\,$\mu$m and 12\,$\mu$m.

We find that the no-track calibrators normally have a lower flux. Furthermore, when we separate the results by the amount of time the fringe spends close to the piezo delay position, we find that observations that spend most of their time near this position have higher fluxes, closer to that of the tracked calibrators, when compared to those that spent very little time near the piezo delay position. However, there is less difference between calibrators which have a moderate amount of frames, $\approx50\%$, recorded close to zero Optical Path Difference (OPD), and those which only have a few frames recorded close to zero OPD.

The discrepancy between tracked and no-track sources could be caused by this difference in average OPD, the distance between the group delay fringe peak and the piezo delay line position, between tracked and no-track sources. A tracked object is normally, on average, 40$\,\mu$m away from the piezo delay. However, a no-track object can spend a significant amount of time close to the 100$\,\mu$m limit. So, even with the limit, this difference may be significant. To test this we set the acceptable OPD range to be more strict. By choosing OPD ranges smaller than the piezo scanning range we should remove this effect at the expense of the number of frames that can be used to calculate the correlated flux. We split the standard OPD range into $20\,\mu$m blocks, setting minimum OPD, minOPD, and maximum OPD, maxOPD, in \textsc{ews} to [0,20,40,60,80] and [20,40,60,80,100] respectively. We compared the uncalibrated correlated flux of the tracked and no-track calibrators for each range. We find that the uncalibrated correlated flux decreases with increasing OPD as expected. However, we also find that even for the narrow OPD ranges the correlated fluxes obtained in tracked and no-track mode do not agree. The difference in correlated fluxes for the two modes shows no consistency or further structure between objects (example in Figure \ref{F: ADU comp}). Therefore, we cannot correct for this difference when reducing and calibrating objects taken with the no-track mode.

Because the conversion factor loosely depends on the average OPD we will still have these correlation losses if we calibrate no-track science with no-track calibrators. Therefore, we include an additional 8\% uncertainty on the correlated flux of all no-track observations at 12\,$\mu$m. This error was derived from the scatter of the conversion factor differences between the comparable tracked and no-track calibrators and can be seen in Figure \ref{F: ADU comp multi calib}. We also show the mean difference for two different OPD ranges, the standard range of 20$\,\mu$m$-100\,\mu$m and a more restricted range of 10$\,\mu$m$-40\,\mu$m to demonstrate that this offers no significant reduction in this error at either wavelength.

\subsubsection{Non-Detections}
\label{Subsection:Non Detections}

Many observations are of insufficient quality to use in fitting. We define an observation as bad if the $9\mu$m atmospheric ozone feature is not clearly present in the uncalibrated correlated flux. However, a true non-detection, one not caused by weather or instrumental failure, can hold important information in interferometry. A low visibility means an object is highly resolved. We made two checks to identify true non-detections.

The first check is using the ESO Observatories Ambient Conditions Database to see if the observation coincided with adverse observing conditions. We also used the changes in the conversion factors of the calibrators \citep{burtscher_observing_2012} to confirm these events.

The second check is to look for correlations between PA, baseline length, and the non-detections. If the non-detections are true then they must either be a low emission period in the AGN or caused by the object being resolved to the interferometer. In the latter case the non-detections would be grouped either above a certain baseline length or at a particular PA if the extended component has an angular dependence. If, on the other hand, other observations with equal or larger baseline lengths at similar PAs have detections then the non-detection is not caused solely by the geometry of the source.

Most of the non-detections were shortly followed by successful no-track observations and, therefore, do not provide any useful upper limits. Fairall\,49 is the only object where we have a true non-detection that was not followed by a no-track observation. However, the upper limit from this measurement provided no further constraint to Fairall\,49, because the measured visibilities have very low uncertainties and are already lower than the limit from the non-detection (see Figure\,\ref{fig:results}). Therefore the upper limit was not included in fitting. This is because it is at a longer baseline length than the other two observations, therefore, it tells us that the visibility does not increase after the 57\,m observation which is already assumed by our simplistic model (see Section \ref{Modelling}).

\subsection{VISIR}
\label{Section:VISIR}
The spectrophotometry from MIDI, used to calculate the visibilities, is usually poor for faint objects \citep{burtscher_observing_2012}. 
Therefore, it is often better to use instead the single-dish spectrophotometry from VISIR as total flux reference. This is done for all objects in this work. Archival VISIR low spectral resolution $N$-band spectra are available for all objects except Fairall\,51 (Fairall\,9 and Fairall\,49 in \citealt{jensen_pah_2017}; MCG-06-30-15, Mrk\,509, NGC\,2110, NGC\,7213 and NGC\,7674 in \citealt{honig_dusty_2010-1}). For Fairall\,51, we obtained a new spectrum with VISIR after its upgrade \citep{kaufl_return_2015, kerber_visir_2016} in October 2018 as part of ESO programme 0101.B-0411 (PI: D. Asmus) in low-resolution mode with standard chopping and parallel nodding, and an on-source exposure time of 5\,min. 
The conditions were reasonably good (water vapour $\sim2.5$\,mm, infrared sky surface temperature $\sim-93\degree$C, optical seeing $\sim0.8$-$1.0$\,arcsec). The data were reduced with the ESO pipeline and flux calibrated using the corresponding acquisition images. In particular, we scale the spectra by the ratio of the synthetic photometry, derived from the acquisition filter transfer function, to the photometry of the acquisition image. The resulting spectra are shown in Figure \ref{fig:Flux comparison}.

In order to check for long-term variability of the AGN at the MIDI wavelengths and better constrain the total flux during the MIDI observations, we obtained two or more epochs of VISIR \textit{imaging photometry} of all objects except Fairall\,49 in 2015 and 2016 (ESO programme IDs 095.B-0376 and 096.B-0369; PI: D. Asmus) in the filters PAH1 (8.59\,$\mu$m) and PAH2\_2 (11.88\,$\mu$m).
Standard imaging mode with chopping and perpendicular nodding with a throw of 10\,arcsec was used.
The on-source exposure times were 7 and 10\,min, respectively, and observations of the science targets were followed by MIR standard stars from the catalogue from \cite{cohen_spectral_1999}, selected to be into a similar direction in the sky.
The data reduction and flux measurements were performed with a custom made \textsc{python} pipeline (Asmus, in prep.).

The resulting fluxes and measured sizes of the nuclei are listed in Table \ref{tab:visir}, whereas the fluxes are averaged over all the epochs, and the FWHM values are the smallest measured for each filter.

\begin{deluxetable*}{l c c c c c}
\tablewidth{0pt}
\tablecaption{VISIR photometry.\label{tab:visir}} 
\tablehead{
\colhead{	Name	}& \colhead{	Dates	}& \colhead{	$F_\nu$(8.6$\,\mu$m)			}& \colhead{	$F_\nu$(12$\,\mu$m)			}& \colhead{	FWHM(8.6$\,\mu$m)	}& \colhead{	FWHM(12$\,\mu$m)	}\\
\colhead{		}& \colhead{	[yy-mm-dd]	}& \colhead{	[mJy]			}& \colhead{	[mJy]			}& \colhead{	[arcsec]	}& \colhead{	[arcsec]	}\\
\colhead{	(1)	}& \colhead{	(2)	}& \colhead{	(3)			}& \colhead{	(4)			}& \colhead{	(5)	}& \colhead{	(6)	}
}
\startdata
Fairall\,9	&	15-07-07/15-08-01/15-08-29	&	195.7	$\pm$	19.6	&	296.0	$\pm$	29.6	&	0.26*	&	0.37*	\\
Fairall\,49	&	10-08-31*	&	314.9*	$\pm$	31.5*	&	523.5*	$\pm$	52.4*	&	0.55*	&	0.73*	\\
Fairall\,51	&	15-07-11/15-07-26/15-07-27/15-09-04	&	223.5	$\pm$	22.3	&	361.2	$\pm$	36.1	&	0.27	&	0.33	\\
MCG-06-30-15	&	15-07-25/15-07-26	&	189.2	$\pm$	18.9	&	334.0	$\pm$	33.4	&	0.29	&	0.35*	\\
Mrk\,509	&	15-07-09/15-07-24/15-07-25	&	161.3	$\pm$	16.1	&	255.1	$\pm$	25.5	&	0.26	&	0.32	\\
NGC\,2110	&	15-11-08/15-12-30/16-02-14/16-02-15/16-02-27	&	201.8	$\pm$	20.2	&	360.9	$\pm$	36.1	&	0.28	&	0.34	\\
NGC\,7213	&	15-07-07/15-07-25	&	70.0	$\pm$	7.2	&	197.0	$\pm$	19.7	&	0.29	&	0.33*	\\
NGC\,7674	&	15-07-08/15-07-16	&	203.8	$\pm$	20.4	&	395.2	$\pm$	39.5	&	0.34*	&	0.40*	\\
\enddata
\tablecomments{
Column descriptions:
(2): Dates of the individual epochs during which always both filters, PAH1 and PAH2\_2 were observed;
(3) and (4): nuclear flux densities at $8.6\,\mu$m and $12\,\mu$m estimated from the filters PAH1 and PAH2\_2, respectively through Gaussian fit photometry. 
The quoted uncertainties are dominated by the systematic uncertainties on the calibrator flux in all cases which is $\le 10$\%;
(5) and (6): the smallest FWHM values in the PAH1 and PAH2\_2 filters, respectively, of all the epochs from Gaussian fitting. 
The major and minor axis are averaged.
Values marked by a * are taken from the VISIR image collection of \cite{asmus_subarcsecond_2014} if they provide better estimates than the new data.
Fairall\,49 is completely taken from the latter work and is listed here for convenience.
}
\end{deluxetable*}

The fluxes of the individual epochs are all consistent with no flux variations during the measurement epoch and also agree within the uncertainties with the historical values and taken from \cite{asmus_subarcsecond_2014}, except for NGC\,7213 and NGC\,2110. 
In the case of NGC\,7213, the flux level in both filters has decreased by $\sim 20\%$ since the last measurement in 2009. 
Thus, we scale the VISIR LR spectrum accordingly to match the 2015 fluxes which are closer to our MIDI observations.
Furthermore, NGC\,2110 showed a flux increase by $\sim 20\%$ over the whole $N$-band between 2007, when the VISIR LR spectrum was taken, to 2010.
From 2010 to 2015, the flux levels seems to have remain constant.
Therefore, we scale the VISIR LR spectrum 20\% up to match the flux levels during the MIDI observations.

Finally, to check if the VISIR data are a good substitute for the MIDI spectrophotometry we compare all the MIDI total flux spectra for each object against its VISIR counterpart (Figure \ref{fig:Flux comparison}). We find good agreement between the spectrophotometries for all objects except Fairall\,49 and NGC\,7213. Both of these have a VISIR spectrophotometry consistently higher than the MIDI observations. The VISIR observations are within errors of the MIDI observations on both these objects but they are consistently above. For Fairall\,49 we only have two observations, therefore, the difference is not significant. Furthermore, the one observation that was significantly lower does not agree with the contemporaneous correlated flux measurement. A comparison between the dimmer MIDI spectrophotometry and its correlated flux counterpart leads to visibilities of greater than 1. Further study of the spectrophotometry shows that the observation is suffering from poor sky subtraction. The remaining spectrophotometry agrees with the VISIR results. We will provide two interpretations of the geometry for NGC\,7213 using both spectrophotometries as the zero baseline value.

\subsection{SEDs}

For each of the objects we attempt to create an infrared spectral energy distribution (SED). To match the angular resolution of VISIR we use archival ISAAC data \citep{moorwood_isaac_1998} (see Table \ref{tab:ISAAC}). There was no ISAAC data available for NGC\,2110, instead NACO was used \citep{lenzen_naos-conica_2003,rousset_naos--first_2003}. The reduction methodology is the same. 

\subsubsection{ISAAC}

The ISAAC data were extracted using the python package \textsc{photutils} \citep{larry_bradley_astropy/photutils:_2019} and \textsc{astropy} \citep{astropy_collaboration_astropy:_2013} with an aperture of radius 0.5\,arcseconds. Each extracted observation was calibrated using the closest standard star in time. The uncertainty on the calibration was found by reducing all observations of the standard star within 1 hour of the selected calibration observation and finding the standard deviation and average count. The error was added in quadrature to the error derived from the background noise. The zero point was calculated for each calibration observation and the average was used as the zero point for the science observation. For the L and M band, chopping and nodding was used. However, it was not used for the H and Ks bands.

\subsubsection{Spitzer}

To cover the longer wavelengths we use archival level 2 Spitzer data (see Table \ref{tab:Spitzer}). All observations of each AGN were collected and binned into 0.1\,$\mu$m bins, the content of these bins were averaged. We then plot the Spitzer and ISAAC data along with the VISIR spectrophotometry in Figure \,\ref{fig:results}. For the sake of clarity, we only plot the low resolution data. These SEDs allow us to look for relationships between SED features and interferometric visibilities (in Section \ref{Subsection:SED results}).

\begin{table}
    \centering
    \caption{Spitzer/IRS observations used in SEDs.}
    \begin{tabular}{l c c c}
    \hline
         Object&Programme ID&Start Date&Mode\\
         \hline \hline
         Fairall\,49&86&2006-04-19&IRS Stare\\
         Fairall\,49&30572&2007-10-05&IRS Stare\\
         Fairall\,51&50588&2008-06-02&IRS Stare\\
         Fairall\,9&86&2003-12-17&IRS Stare\\
         Fairall\,9&30572&2007-06-13&IRS Stare\\
         Fairall\,9&526&2009-01-20&IRS Stare\\
         MCG-06-30-15&86&2004-06-28&IRS Stare\\
         MCG-06-30-15&3269&2005-02-15&IRS Map\\
         MCG-06-30-15&30572&2007-07-29&IRS Stare\\
         Mrk\,509&86&2004-05-14&IRS Stare\\
         Mrk\,509&30572&2006-11-19&IRS Stare\\
         NGC\,2110&86&2004-03-22&IRS Stare\\
         NGC\,2110&30572&2007-11-04&IRS Stare\\
         NGC\,7213&86&2004-05-15&IRS Stare\\
         NGC\,7213&3269&2005-05-25&IRS Map\\
         NGC\,7213&30572&2007-06-13&IRS Stare\\
         NGC\,7674&3269&2004-12-10&IRS Map\\
         NGC\,7674&3605&2005-07-05&IRS Stare\\
    \end{tabular}
    
    \label{tab:Spitzer}
\end{table}

\section{Modelling}
\label{Modelling}


As an initial check for position angle dependence in the MIDI data we use the same method demonstrated for ESO\,323-G77 in \citet{leftley_new_2018}. To summarise this method, we compare the visibility at 12\,$\mu$m\,$\pm\,0.4\,\mu$m to the baseline length and separate the observations by PA into three groups consisting of $0^\circ\leqslant$\,PA\,$<60^\circ$, $60^\circ\leqslant$\,PA\,$<120^\circ$, and $120^\circ\leqslant$\,PA\,$<180^\circ$. For all further discussion we only use the visibilities at 12.0$\pm$0.4$\mu$m. We only report at 12\,$\mu$m to be consistent with \citet{burtscher_diversity_2013,lopez-gonzaga_mid-infrared_2016} and because we are less effected by the no track mode error as well as other calibration losses. Fairall\,49 is the only object to show a significant change in visibility with baseline length in the initial check, though we only have two observations from which to draw this conclusion. Since we do not find PA dependent structure in any objects we rotate the bins incrementally by $10^\circ$ through $50^\circ$ to rule out that the lack of structure was due to bin choice. We do not find a PA dependence in any objects, therefore, we can simplify our model for these AGN.

Because our objects show no clear variation in visibility with PA, at MIDI's level of accuracy, we fit a constant plus a radially symmetric Gaussian. The constant represents the fraction of the total flux that is unresolved to MIDI at all available baseline lengths. The fitting method is the same as detailed in \citet{leftley_new_2018} with two free variables, the FWHM of the Gaussian ($\theta$) and the unresolved flux fraction (p$_f$). In most cases the full width at half maximum (FWHM) of the extended component is unbound when fit.

In the cases where only a constant can be reliably fit the Gaussian can give us a lower bound for the radial size of the extended dust. For these objects we perform two fits to the data. In the second fit we additionally apply a non flat prior in the $\Theta$ direction where $\Theta$ is the FWHM of the Gaussian. This prior is linear and is defined as $-0.15\Theta$. This makes smaller values for $\Theta$ more likely without affecting the result of the fit itself. After testing this prior against real data we find that any loss in accuracy caused by the introduced bias is far less than the derived error of the fit. This second fit provides us with our lower limit only, the point source fraction is still derived from the original flat prior fit.

\subsection{SED modelling}
The SED of an object can provide zeroth order information about the distribution of dust in the object. This is because the IR emission has a contribution from thermally emitting dust and the dust is heated by a central source. In the SED this manifests as two distinct bumps, the $3-5\,\mu$m bump and the $20\,\mu$m bump \citep{edelson_spectral_1986,elvis_atlas_1994,mullaney_defining_2011}. These bumps correspond to two dusty components with temperatures of $\sim700$\,K and $\sim200$\,K respectively. Therefore, the SED may be able to predict the unresolved fraction seen by the VLTI. This would be a huge advantage when finding candidates for observations. We hypothesise that a black body fit to the warm or hot dust bumps in the SED of an AGN will be able to tell us the fraction of the $12\,\mu$m emission that the wind and disk, respectively, are responsible for. This would then translate into an unresolved source fraction and, therefore, a minimum visibility for future observations with, e.g., VLTI/MATISSE.

For this purpose we fit two simple black bodies to the SED and compare the $12\,\mu$m emission fraction to the unresolved source fraction from MIDI.

\section{Results}
\label{S:Results}

\begin{table}

\caption{Modelling results}
\begin{tabular*}{0.47\textwidth}{l @{\extracolsep{\fill}} c c c}\hline
Object&$\Theta$ (mas)&$\Theta_{up}$ (mas)&$\mathrm{p}_f$\\ \hline \hline
Fairall\,51&$\geqslant65$&330&$0.54^{0.01}_{-0.01}$\\
Fairall\,49&$38.28^{3.31}_{-2.80}$&730&$0.51^{0.03}_{-0.03}$\\
Fairall\,9&$\geqslant22$&370&$0.70^{0.05}_{-0.05}$\\
MCG-06-30-15&$\geqslant43$&350&$0.43^{0.02}_{-0.02}$\\
NGC\,2110&$\geqslant21$&340&$0.53^{0.04}_{-0.06}$\\
Mrk\,509&$18.13^{6.83}_{-4.02}$&320&$0.35^{0.16}_{-0.21}$\\
NGC\,7213&$\geqslant20$&330&$0.63^{0.04}_{-0.05}$\\
NGC\,7674&$\geqslant64$&400&$0.29^{0.04}_{-0.04}$\\
\end{tabular*}
\label{Table:results}
\tablecomments{$\Theta$ is the FWHM of the extended Gaussian, p$_f$ is the unresolved source fraction, and $\Theta_{up}$ is the upper limit of the FWHM estimated from the single-dish FWHM of the nucleus \citep[][\url{http://dc.g-vo.org/sasmirala}]{asmus_subarcsecond_2014}}.
\end{table}

\subsection{General Results}

We find that all of the studied sources are partially resolved. However, only two show any clear, reliable change in visibility with baseline length, at MIDI's level of accuracy, when we perform our modelling. For the unconstrained objects we can only deduce lower limits to the angular sizes of the resolved dust as well as the unresolved source fraction.

Fairall\,49 and Mrk\,509 show an extended dust feature when fit with the combination of a Gaussian and unresolved source. However, both objects have too scarce a sampling of the \textit{uv} plane to say if the Gaussian and unresolved source model is an accurate descriptor of their geometry. This model has well described all 25 previous AGN from literature. Of these 25 objects, 13 had both components, 7 had the extended component over-resolved and could only constrain the unresolved fraction, 4 only showed the extended component (marginally resolved), and 1 was unresolved \citep{burtscher_diversity_2013,lopez-gonzaga_mid-infrared_2016, fernandez-ontiveros_embedded_2018,fernandez-ontiveros_compact_2019}. Therefore, we take this model as a general description of AGN in the mid-IR. 

\subsection{Results for Individual Objects}

The plotted results for each object can be found, with the \textit{uv} plane and SED, in Figure \ref{fig:results}. The probability distribution uses the flat prior for $\Theta$ and the model in the visibility plot shows the lower limit of the FWHM, as determined by the non flat prior, where the FWHM could not be constrained. Table \ref{Table:results} contains the results for the fitted model. The distances given are the NED, CMB, Hubble distance at the time of writing (H$_0= 67.8$\,km/sec/Mpc, $\Omega$matter\,$ = 0.308$, $\Omega$vacuum\,$= 0.692$).

\textbf{Fairall\,9} is at a distance of 206\,Mpc. It hosts a Sy1.2 nucleus that has three observations which show no clear change in visibility with BL. Therefore, we tentatively report that this objects extended component is over-resolved and the unresolved source fraction is responsible for 68\% of the flux.

\textbf{Fairall\,49} hosts a Sy2 nucleus and is at a distance of 88\,Mpc. It has two MIDI observations ,separated by one day, of high enough S/N to draw direct conclusions from. The $\sim3$\,M$\lambda$ baseline observation is unusually high when compared to the average MIDI single-dish spectrophotometry. At longer wavelengths it has a visibility of greater than 1. If we do not average the MIDI spectrophotometry and compare the unusually high observation to its contemporaneous spectrophotometry only we still get visibilities of greater than 1. However, when we compare the correlated flux to the VISIR spectrum, or the MIDI spectrophotometry of the second observation, we find that this observation gives sensible visibilities. Further inspection shows that the MIDI spectrophotometry may be suffering from poor sky subtraction. We therefore deem the interferometric observation reliable. When fit with the model, we constrain an extended component of 38\,mas and unresolved source of 0.51. The non-detection, Section \ref{Subsection:Non Detections}, tells us the visibility is unlikely to increase at longer baselines.

\textbf{Fairall\,51} has a Sy1 nucleus and is at a distance of 62\,Mpc. It is partially resolved with an extended component that is larger than observable with MIDI. We conclude that Fairall\,51 shows no change in visibility with baseline length or PA at MIDI's level of accuracy and available baselines. It has an unresolved fraction of 0.54 and the FWHM of the resolved component is between 65\,mas and 330\,mas. However, while Fairall\,51 is fully consistent with an unresolved source it the data suggests a bump in the visibility between 7\,M$\lambda$ and 10\,M$\lambda$ (Figure \ref{fig:results}). This bump is well described by a simple dust ring around the unresolved source. Further observations will be necessary to conclude if this bump is real.

\textbf{MCG-06-30-15} is host to a Sy1 nucleus at a distance of 38.3\,Mpc. It shows partially resolved structure that has an unresolved fraction of 0.43, however, the extended structure of this object is over-resolved. The lower limit on the FWHM of the extended dust is 43\,mas with an upper limit of 350\,mas.

\textbf{MRK\,509} hosts a Sy1.5 nucleus and is at a distance of 148\,Mpc. It has three observations that show the object is partially resolved and suggests an extended dust component. The extended component has a FWHM of 18\,mas, the unresolved fraction, however, is essentially unconstrained with an upper limit of 0.5.

\textbf{NGC\,2110} contains a Sy2 nucleus at a distance of 35.5\,Mpc. It only has visibility measurements for a small range of baseline lengths. These points are mostly in agreement and show no angular dependant structure. Because of the small range of baselines we cannot say if we see baseline length dependant structure. Therefore, we cannot deduce the unresolved fraction or radial size. We find an upper limit of the unresolved source fraction of 0.6.

\textbf{NGC\,7213} hosts a Sy1.5 nucleus at a distance of 22.8\,Mpc. It is partially resolved and shows an unresolved fraction of 0.63. The lower limit of the FWHM is 20\,mas and the upper limit is 330\,mas. However, in this object we find that the MIDI spectrophotometry is much dimmer than the VISIR spectrophotometry. We compare the correlated flux data with the MIDI spectrophotometry and find that this gives unphysical, high visibilities of >1. Even using solely the brightest MIDI spectrum we still find the object is completely unresolved at all baseline lengths. Investigation into the MIDI spectrophotometry shows that it suffers from poor sky subtraction, possibly due to the strong, variable background described in \citet{burtscher_observing_2012}. Therefore, out of the two conclusions given for this object we use the result derived using the VISIR spectrum.

\textbf{NGC\,7674} has a Sy2 nucleus and is at a distance of 122\,Mpc. It has only two observations covering a large range of baselines. We can reliably say that it is partially resolved and it is likely that it has an over-resolved extended component and an unresolved fraction of $\approx0.29$. The lower limit for the extended component is 64\,mas. When viewed with VISIR, the object shows an extended dust feature with a major axis FWHM of 400\,mas \citep{asmus_subarcsecond_2016}.

\subsection{SED visibility fractions}
\label{Subsection:SED results}
When we fit a simplistic black body to the hot dust feature we find no significant correlation between the unresolved source fraction and the fraction of the flux that the hot dust is responsible for. We find that the hot dust in these objects is best explained by a temperature of either 750\,K or 500\,K. We also do not find a correlation within each of these temperatures. This could be due to small number statistics so we cannot rule out the intrinsic presence of a correlation.

As a further test we fit a black body to the cooler dust component, best fit by a temperature of $\sim150$\,K, and found no correlation between between the unresolved source fraction and the warm dust 12\,$\mu$m flux contribution fraction.

\begin{figure*}[b]
    \centering
    \includegraphics[width=0.24\textwidth,trim={0cm 0cm 0.5cm 0cm},clip]{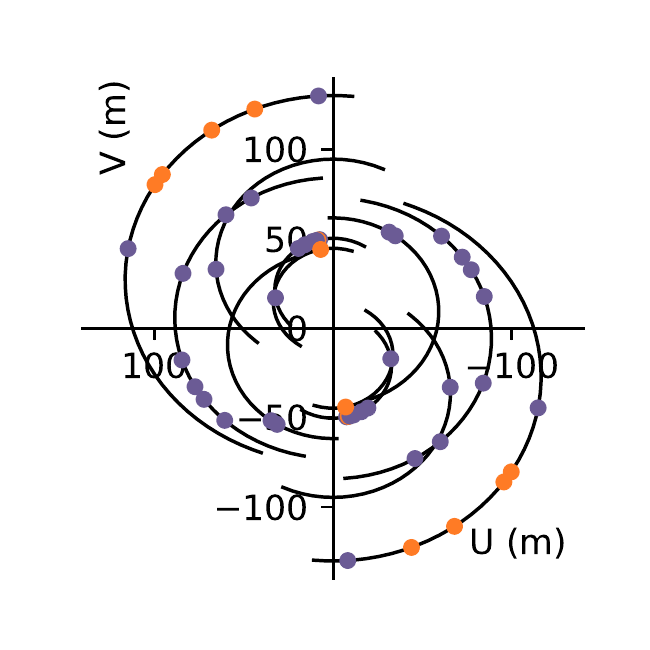}
    \includegraphics[width=0.24\textwidth,trim={0cm 0cm 0.5cm 0cm},clip]{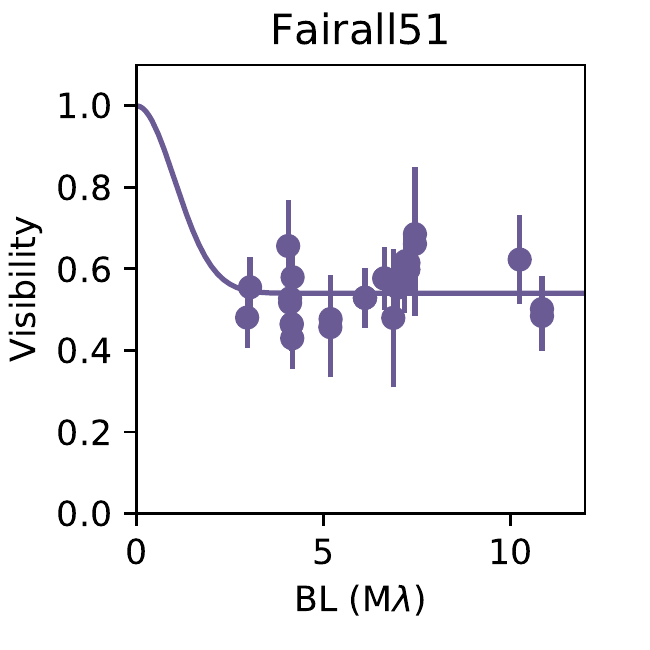}
    \includegraphics[width=0.26\textwidth,trim={0cm 0.4cm 0.44cm 0cm},clip]{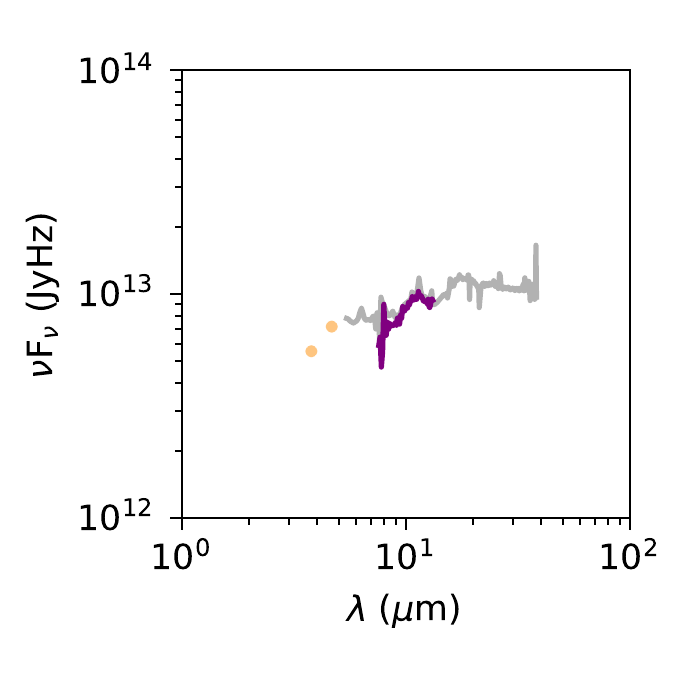}
    \includegraphics[width=0.24\textwidth,trim={0cm 0cm 0.5cm 0.5cm},clip]{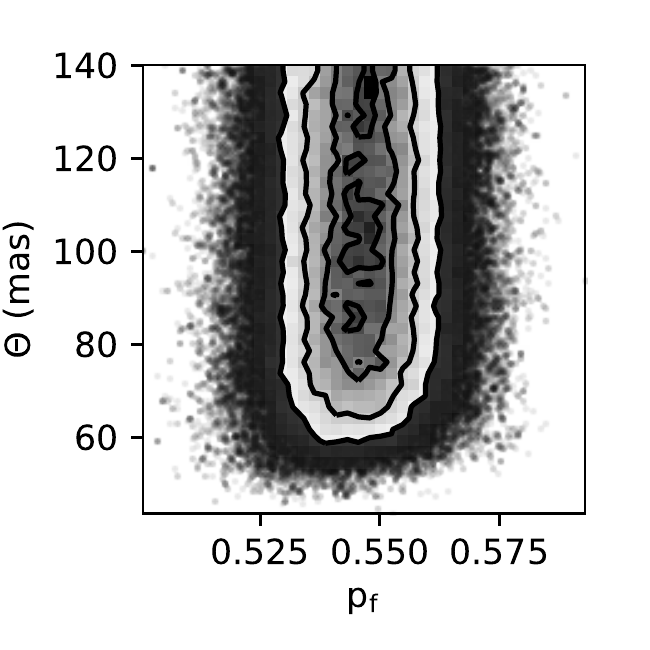}

    \centering
    \includegraphics[width=0.24\textwidth,trim={0cm 0cm 0.5cm 0cm},clip]{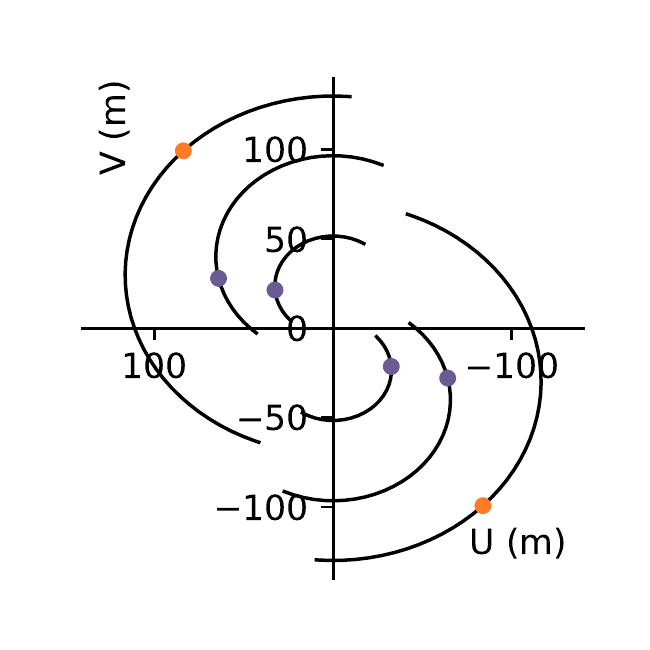}
    \includegraphics[width=0.24\textwidth,trim={0cm 0cm 0.5cm 0cm},clip]{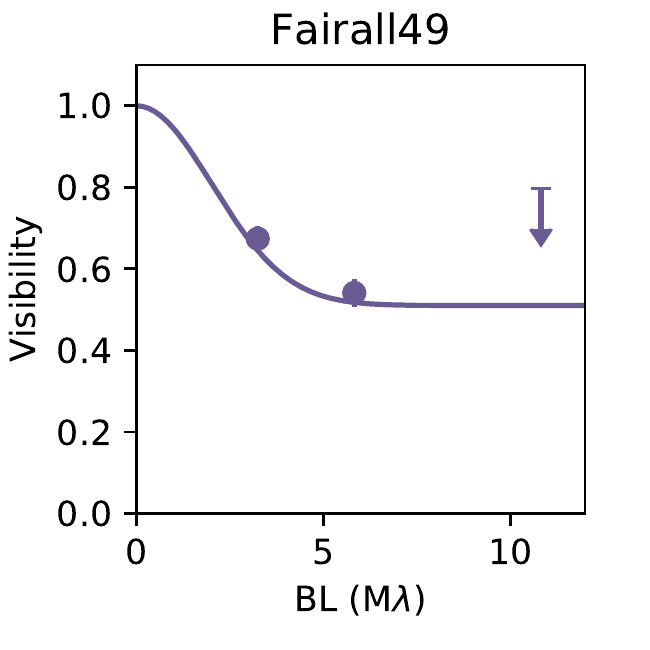}
    \includegraphics[width=0.26\textwidth,trim={0cm 0.4cm 0.44cm 0cm},clip]{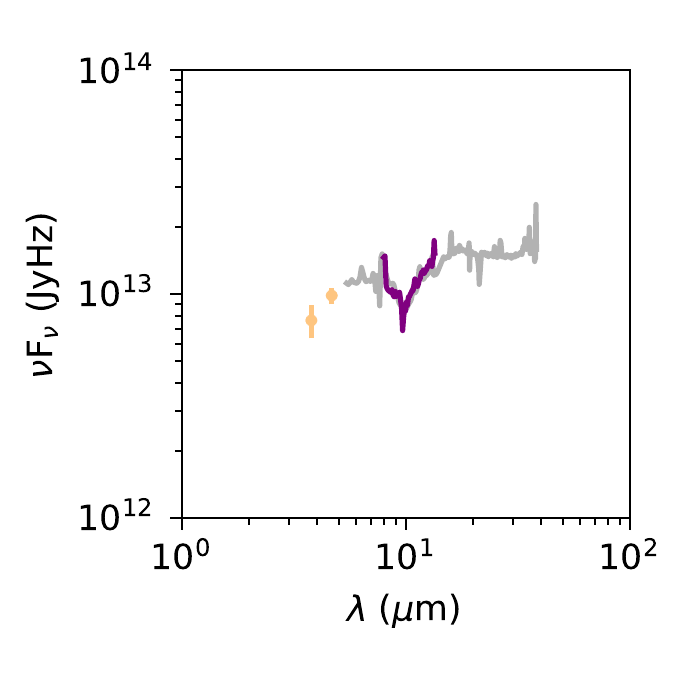}
    \includegraphics[width=0.24\textwidth,trim={0cm 0cm 0.5cm 0.5cm},clip]{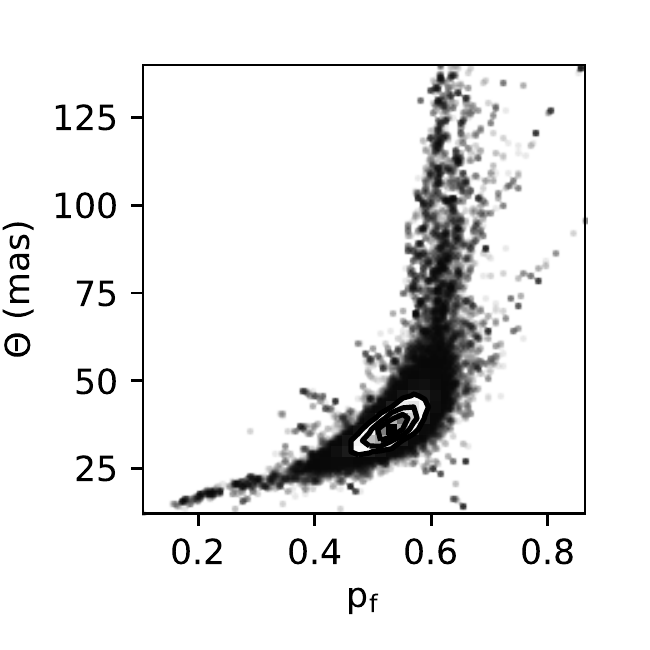}
    
    \centering
    \includegraphics[width=0.24\textwidth,trim={0cm 0cm 0.5cm 0cm},clip]{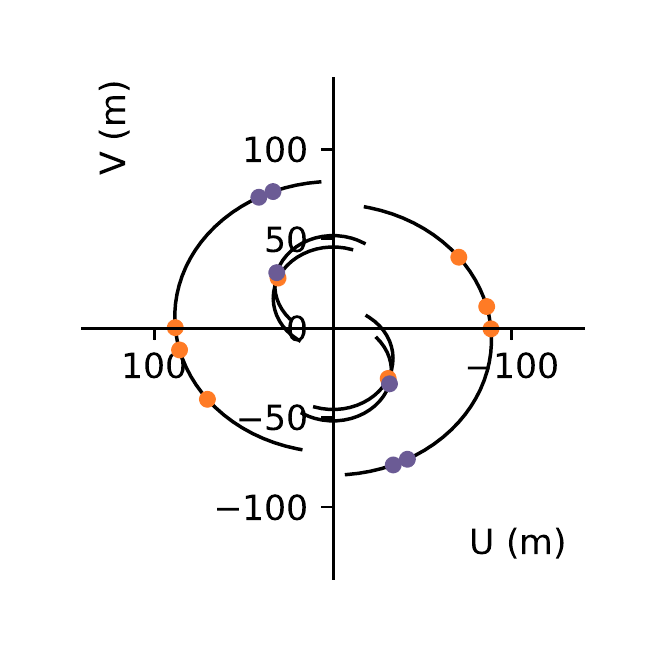}
    \includegraphics[width=0.24\textwidth,trim={0cm 0cm 0.5cm 0cm},clip]{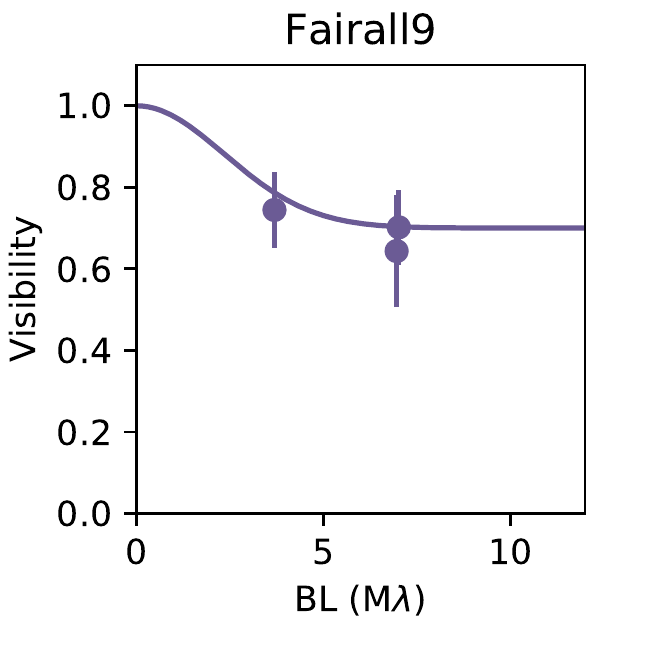}
    \includegraphics[width=0.26\textwidth,trim={0cm 0.4cm 0.44cm 0cm},clip]{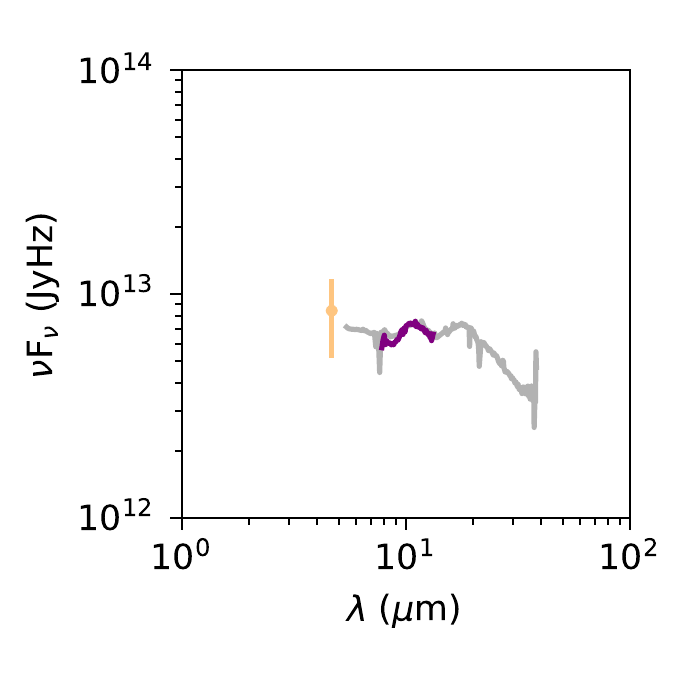}
    \includegraphics[width=0.24\textwidth,trim={0cm 0cm 0.5cm 0.5cm},clip]{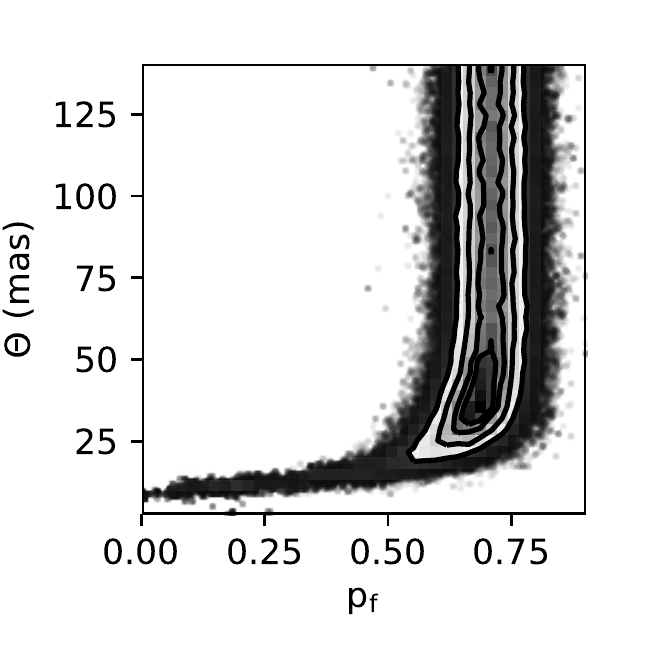}

    \centering
    \includegraphics[width=0.24\textwidth,trim={0cm 0cm 0.5cm 0cm},clip]{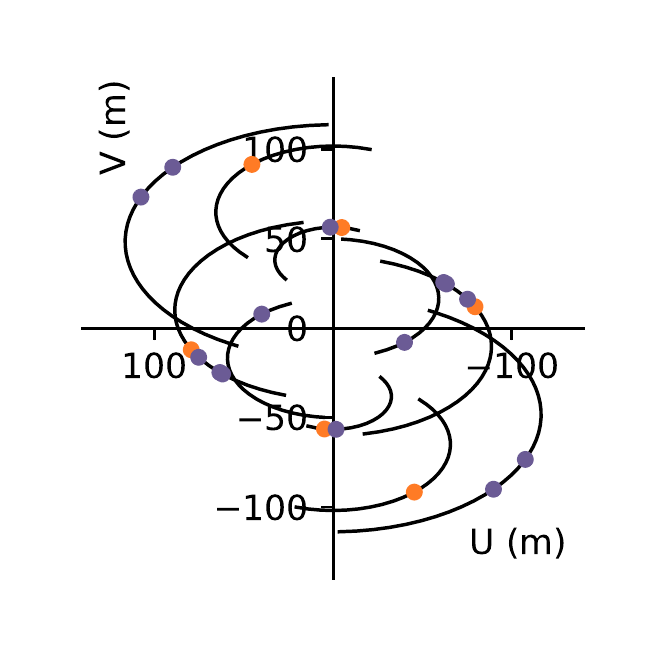}
    \includegraphics[width=0.24\textwidth,trim={0cm 0cm 0.5cm 0cm},clip]{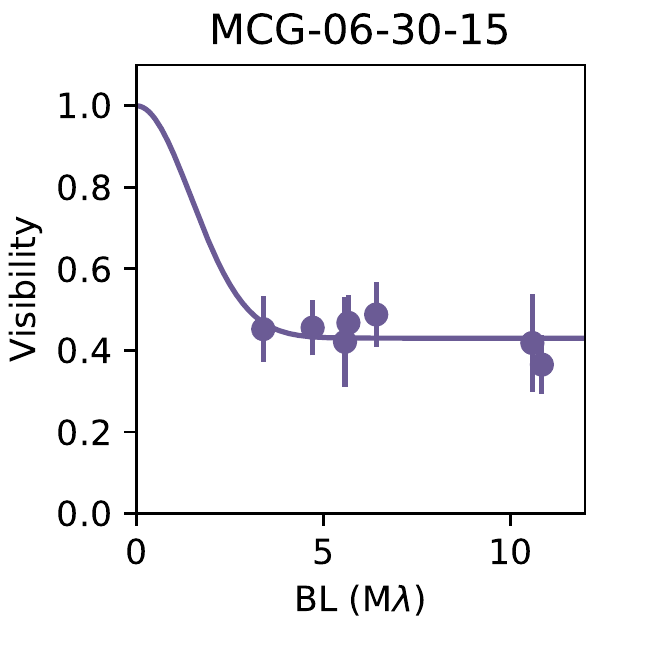}
    \includegraphics[width=0.26\textwidth,trim={0cm 0.4cm 0.44cm 0cm},clip]{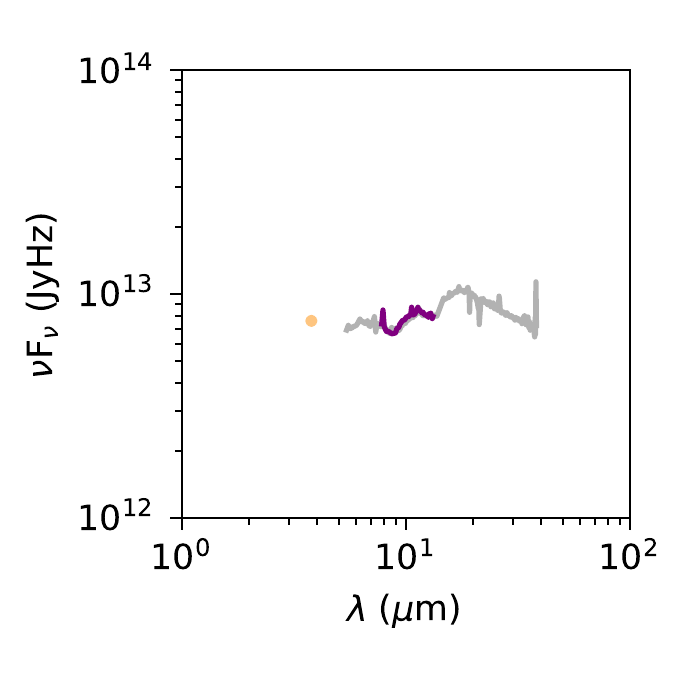}
    \includegraphics[width=0.24\textwidth,trim={0cm 0cm 0.5cm 0.5cm},clip]{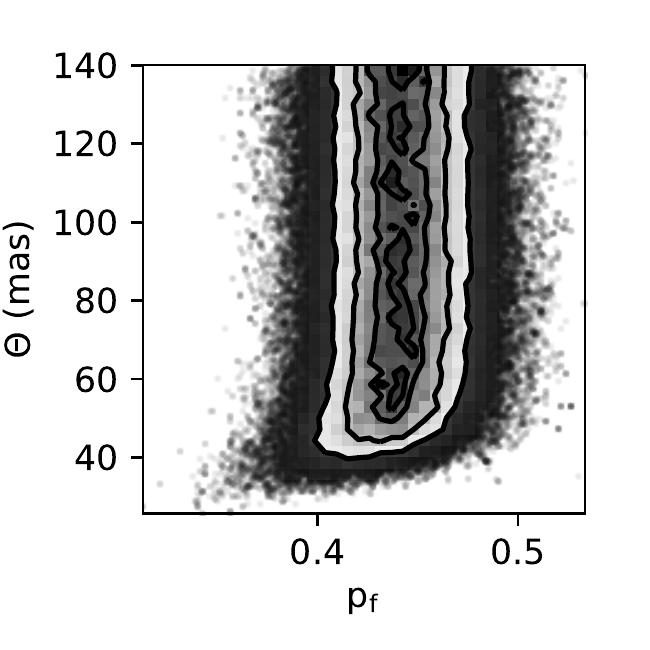}

    \caption{The subplots show, from left to right, the \textit{uv} plane, the visibility measurements and best model fit, the SED, and the geometric model unresolved source fraction vs. extended component FWHM PDF. Each row shows the result for one object. In the \textit{uv} plane, the purple points represent observations of high enough quality to be used directly in our analysis, the orange points are the excluded observations. The visibility plot depicts the visibility of each "good" observation vs. baseline length with the best fit geometric model overlaid. The SED contains data from ISAAC (pale orange), VISIR (purple), and Spitzer (grey).}
\end{figure*}

\begin{figure*}
    \ContinuedFloat
    \centering
    \includegraphics[width=0.24\textwidth,trim={0cm 0cm 0.5cm 0cm},clip]{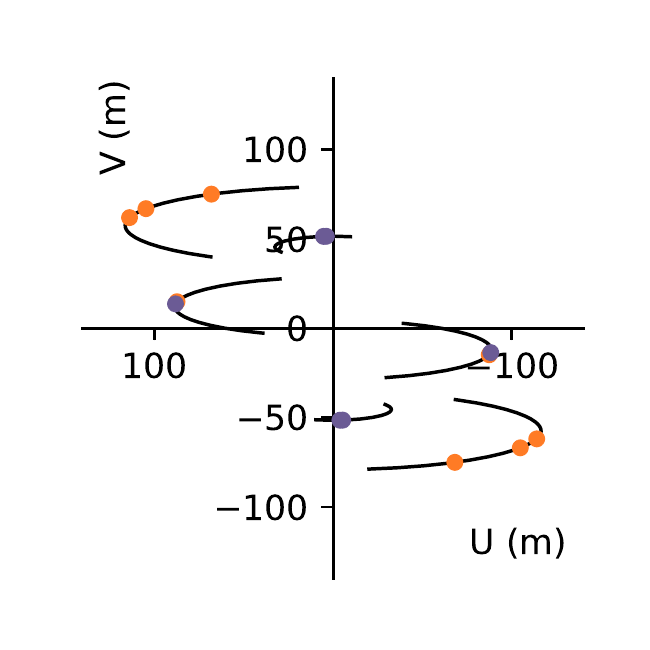}
    \includegraphics[width=0.24\textwidth,trim={0cm 0cm 0.5cm 0cm},clip]{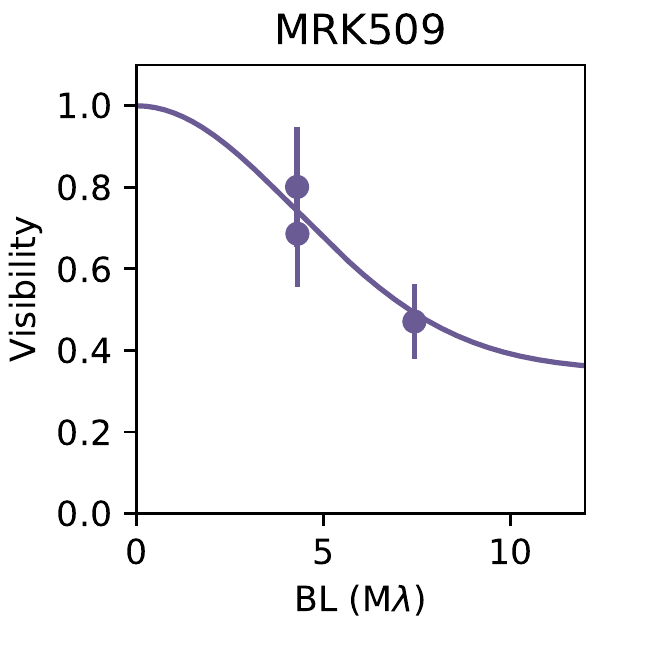}
    \includegraphics[width=0.26\textwidth,trim={0cm 0.4cm 0.44cm 0cm},clip]{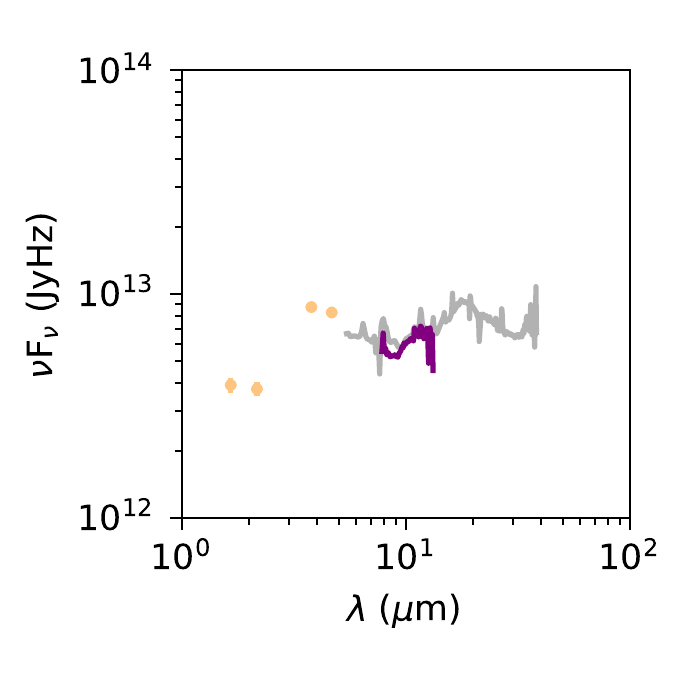}
    \includegraphics[width=0.24\textwidth,trim={0.0cm 0.0cm 0.5cm 0.5cm},clip]{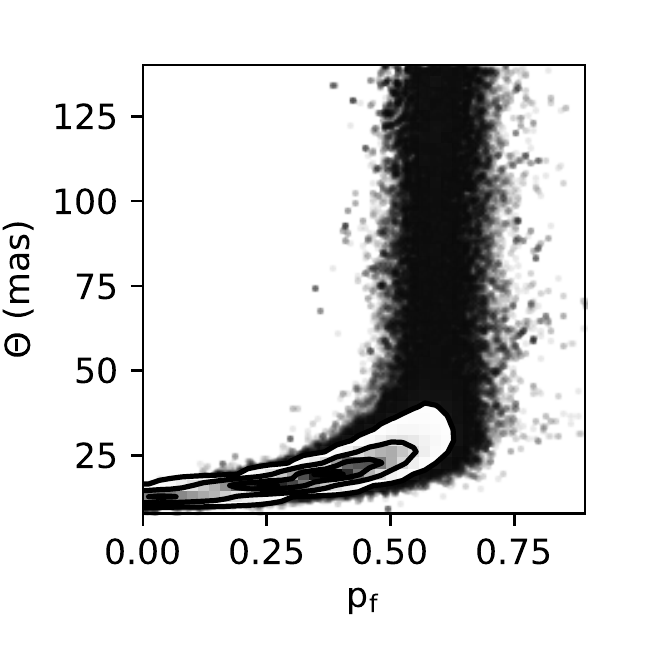}
    
    \centering
    \includegraphics[width=0.24\textwidth,trim={0cm 0cm 0.5cm 0cm},clip]{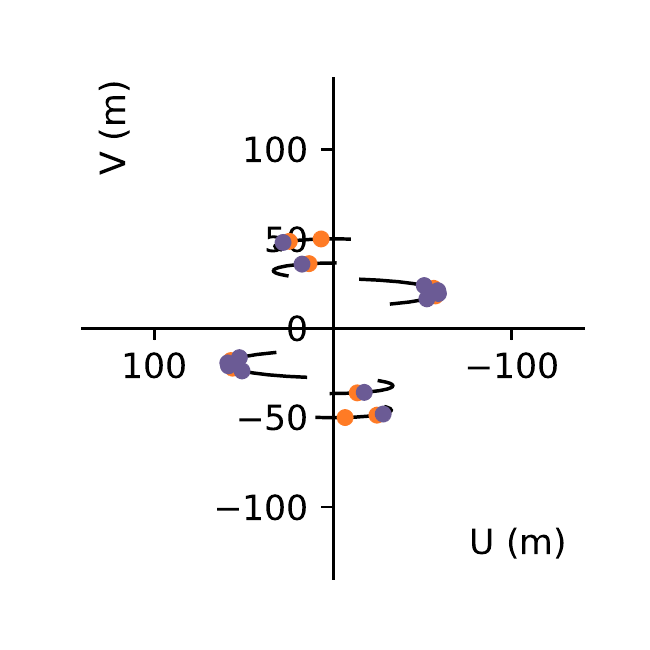}
    \includegraphics[width=0.24\textwidth,trim={0cm 0cm 0.5cm 0cm},clip]{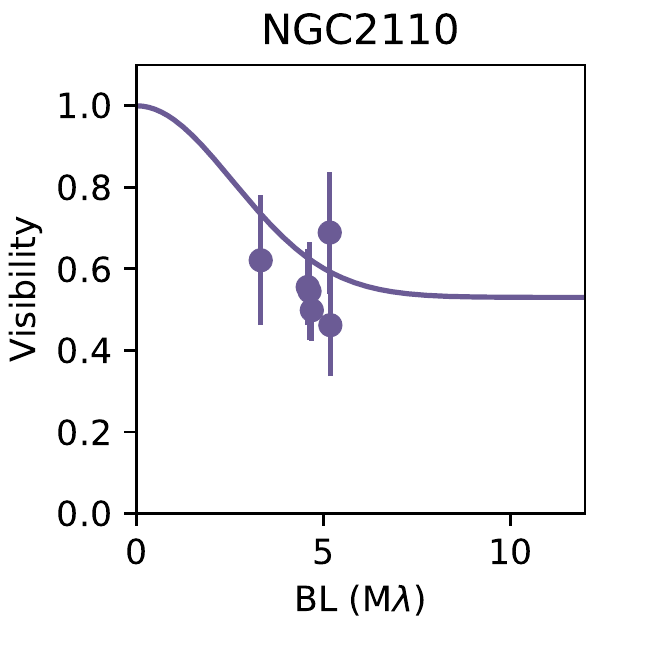}
    \includegraphics[width=0.26\textwidth,trim={0cm 0.4cm 0.44cm 0cm},clip]{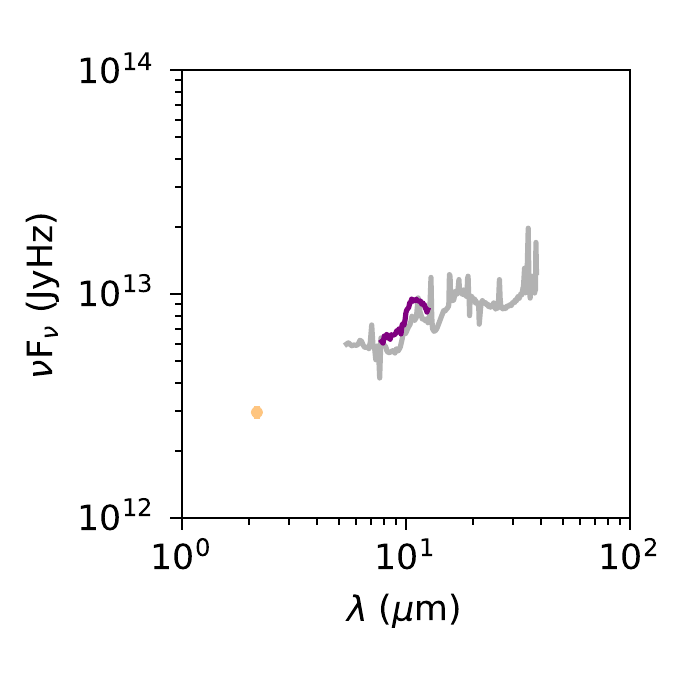}
    \includegraphics[width=0.24\textwidth,trim={0cm 0cm 0.5cm 0.5cm},clip]{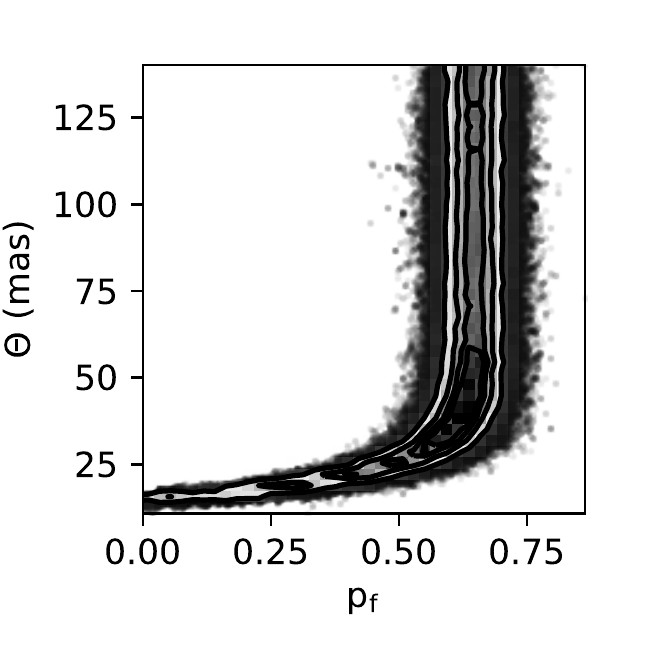}

    \centering
    \includegraphics[width=0.24\textwidth,trim={0cm 0cm 0.5cm 0cm},clip]{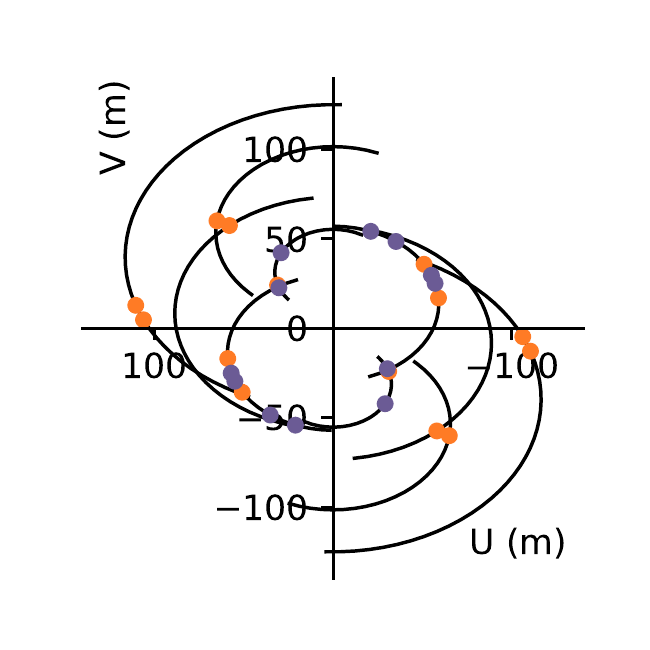}
    \includegraphics[width=0.24\textwidth,trim={0cm 0cm 0.5cm 0cm},clip]{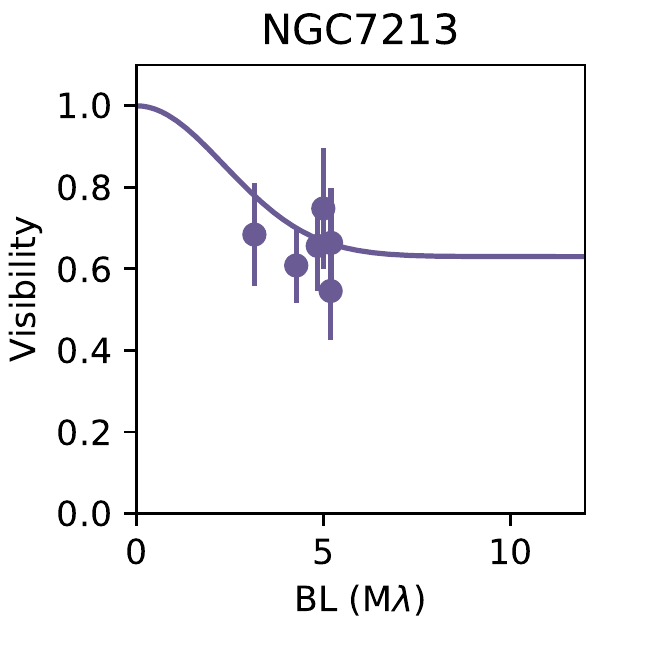}
    \includegraphics[width=0.26\textwidth,trim={0cm 0.4cm 0.44cm 0cm},clip]{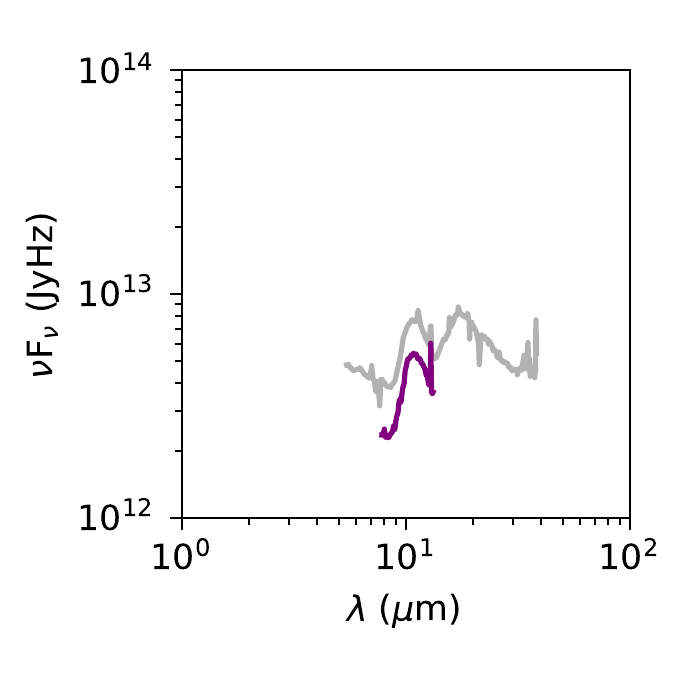}
    \includegraphics[width=0.24\textwidth,trim={0cm 0cm 0.5cm 0.5cm},clip]{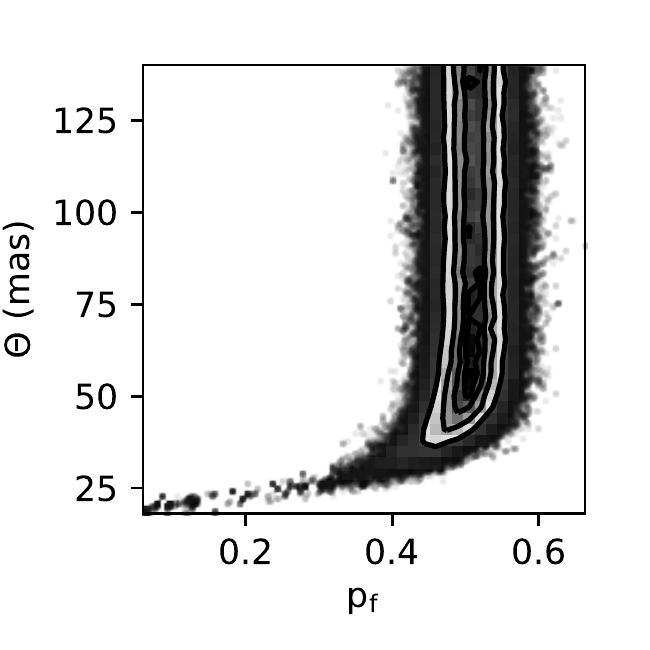}

    \centering
    \includegraphics[width=0.24\textwidth,trim={0cm 0cm 0.5cm 0cm},clip]{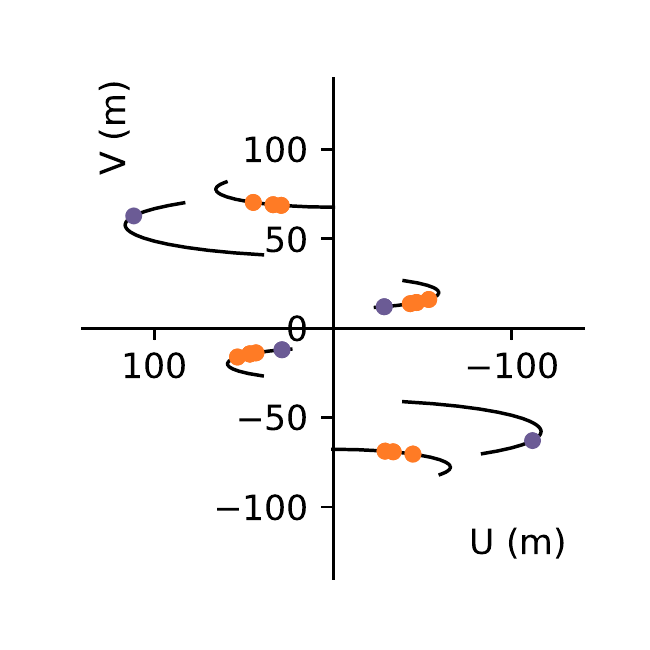}
    \includegraphics[width=0.24\textwidth,trim={0cm 0cm 0.5cm 0cm},clip]{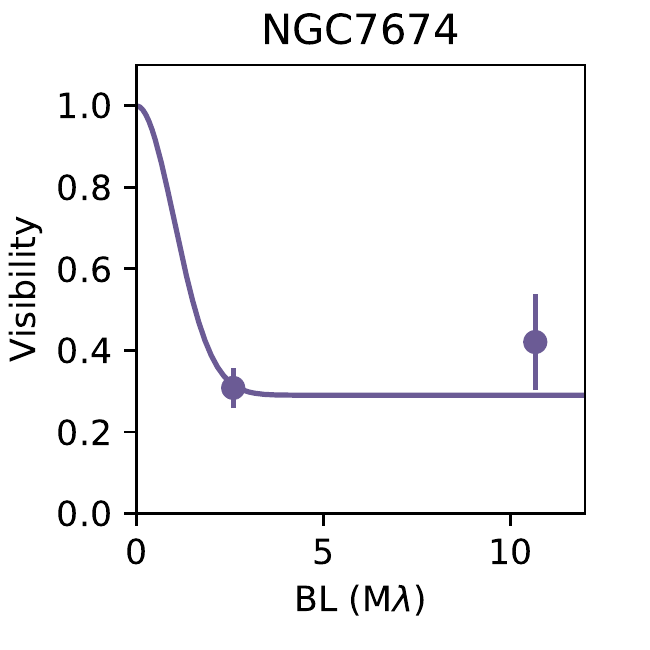}
    \includegraphics[width=0.26\textwidth,trim={0cm 0.4cm 0.44cm 0cm},clip]{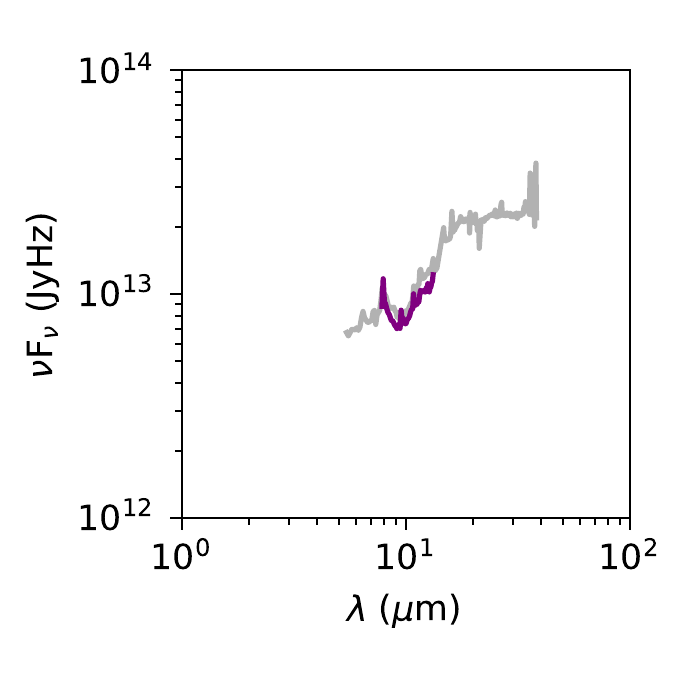}
    \includegraphics[width=0.24\textwidth,trim={0cm 0cm 0.5cm 0.5cm},clip]{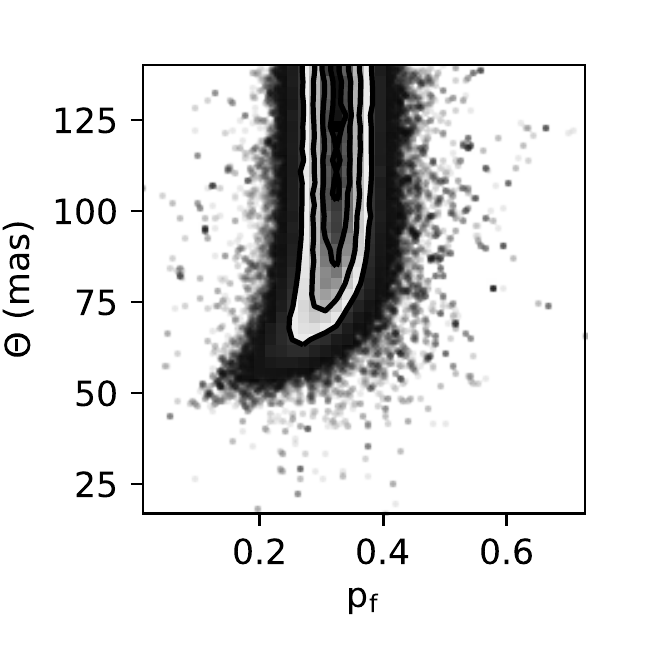}

    \caption{Cont.}
    \label{fig:results}
\end{figure*}

\subsection{Results on the correlation between extended flux fraction and Eddington ratio}
\label{Subsection:eddratres}

\citet{leftley_new_2018} showed that the unresolved component was dominated by the hot dust in ESO\,323-G77. When the source was compared to NGC\,3783, it was thought that there could be an evolution of dust distribution with Eddington ratio. Here, we test this hypothesis further. For our 8 objects, we have the flux of the unresolved and resolved components, we can find the flux values for 24 more objects from previous work \citep{burtscher_diversity_2013,lopez-gonzaga_mid-infrared_2016,fernandez-ontiveros_embedded_2018} and we use the values for ESO\,323-G77 from \citet{leftley_new_2018}. As an initial check we compare the resolved and unresolved luminosities to a proxy for Eddington ratio. Our proxy for Eddington ratio is the ratio of intrinsic X-ray luminosity collected from \citet{asmus_subarcsecond_2015} and the black hole mass taken from various works in the literature \citep{onken_mass_2002,greenhill_warped_2003,lodato_non-keplerian_2003,peterson_central_2004,garcia-rissmann_atlas_2005,denney_mass_2006,gu_emission-line_2006,wang_unified_2007,cappellari_mass_2009,ho_search_2009,makarov_hyperleda._2014} (see Table \ref{tab:Summary} for a summary). To convert intrinsic X-ray luminosity to bolometric we use a conversion factor of 10 \citep{vasudevan_power_2010}.

The black hole mass is a major source of uncertainty in the Eddington ratio. We use maser emission and reverberation mapping where available, which have a typical uncertainty of <0.1\,dex. However, our masses are primarily derived from stellar velocity distribution (SVD). For SVDs we use the intrinsic M-$\sigma$ scaling relation from \citet{shankar_selection_2016} with a scatter of 0.25\,dex. If none of these mass determinations are available, we use H$\beta$ and \ion{O}{3} which have uncertainties of $\geqslant$0.5\,dex.

\begin{figure*}
    \centering
    \includegraphics[width=\textwidth]{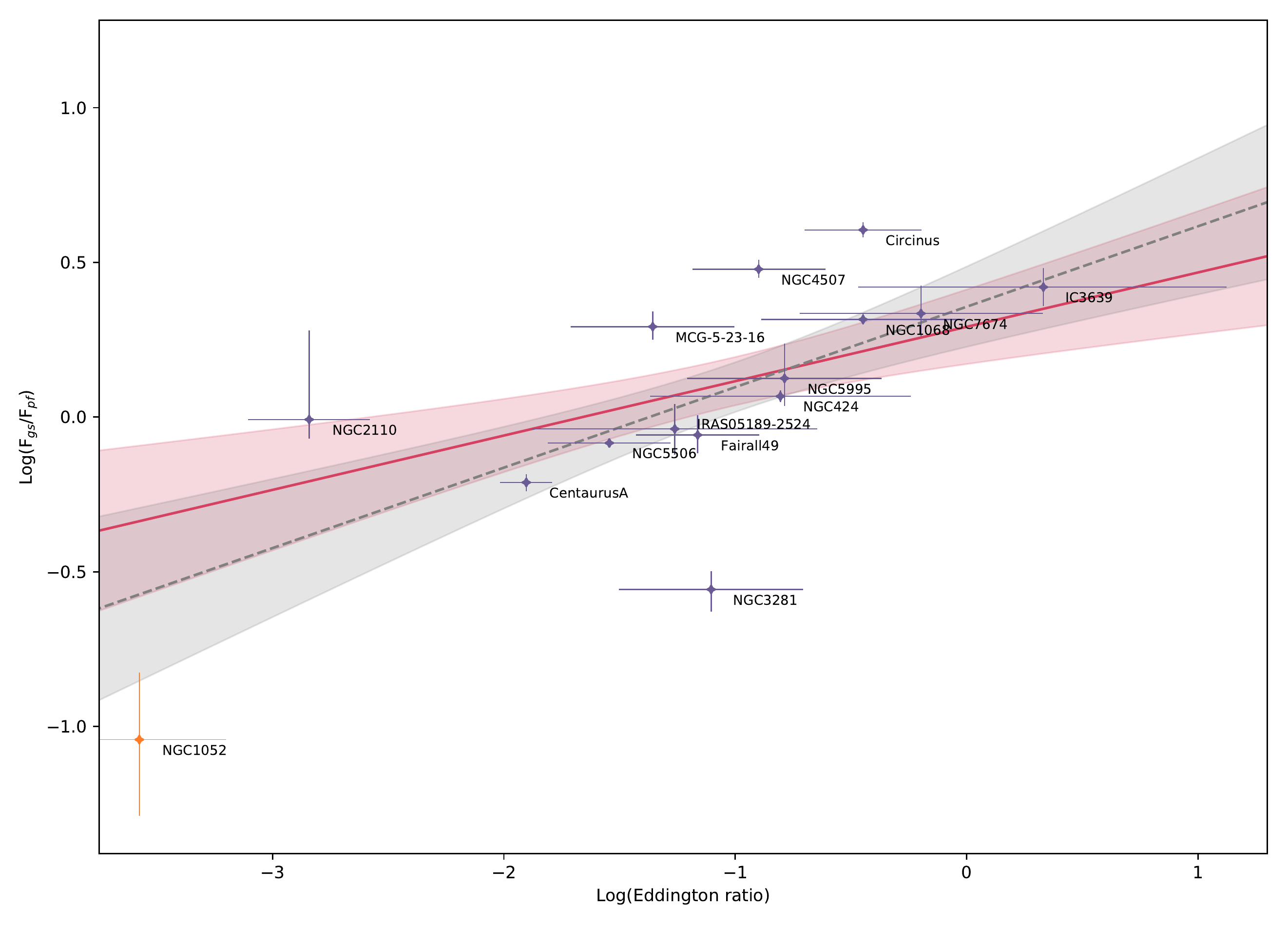}
    \caption{The of ratio extended source flux to unresolved source flux for Sy2 objects compared to their Eddington ratio. The purple points are the objects discussed in this paper, the orange point is NGC\,1052 from \citet{fernandez-ontiveros_compact_2019}. The red line overlaid is the linear best fit, with $1\sigma$ errors, to the purple points and the grey dashed line is the same but includes the orange point.}
    \label{fig:Eddrat}
\end{figure*}

\begin{figure}
    \centering
    \includegraphics[width=0.5\textwidth]{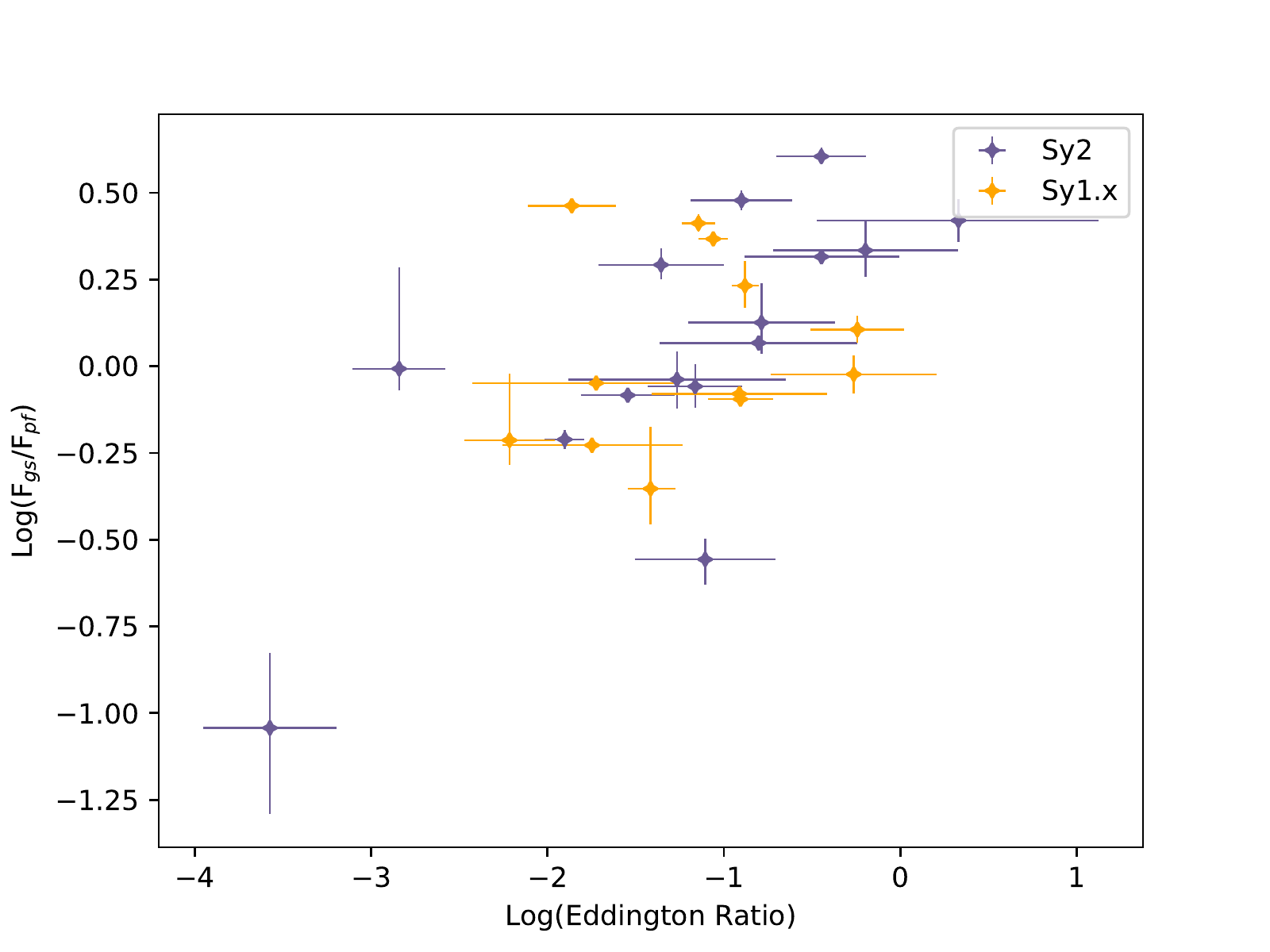}
    \caption{Flux ratio against Eddington ratio for all objects, including NGC\,1052.}
    \label{fig:Eddratall}
\end{figure}

We found no correlation between the point source flux, or extended source flux, and Eddington ratio using the Spearman rank. However, we do find a tentative correlation between the ratio of resolved (F$_{\mathrm{gs}}$) and unresolved (F$_{\mathrm{pf}}$) flux with Eddington ratio for Sy2 AGN (Figure \ref{fig:Eddrat}). In Sy1s we see a large scatter and no correlation (see Figure \ref{fig:Eddratall}). Although, the Sy1s are consistent with the correlation found in Sy2s, however, the Sy1s cover a smaller range of Eddington Ratios. If we take the Sy2 data without errors we get a z (standard score) value of $2.8$ when testing the null hypothesis which, in this case, is that the data are uncorrelated. Equivalently this is a p-value of 0.005 where p is the the two sided Spearman rank null hypothesis probability. As a more rigorous test of this correlation we Bootstrap, with replacement, the Spearman rank so as to include the errors on the measurements and to reduce the effect of outliers on any correlation.

Instead of just taking the $1\sigma$ errors on the objects in this paper and the upper limit of NGC\,2110 we can directly use their probability distributions from the MCMC Bayesian fitting which is a better representation of their true probability distributions. When we do this for all our objects the z value becomes $2.4$ (p-value of 0.016). The Spearman rank correlation coefficient (rho) using the errors is 0.65.

In these results we do not include NGC\,1052 from \citet{fernandez-ontiveros_compact_2019} due to their conclusion that the mid-IR photons do not arise from thermal emission. This would make it unsuitable for our study of radiation pressure. However, the presence of silicate emission and nuclear obscuration suggests thermally emitting dust which could mean a very compact dust core is responsible for significant parts of the mid-IR emission. Therefore, we provide results with and without this object. With this object the z value without errors becomes $3.3$ and with bootstrapped errors $3.0$ (p-value of 0.001 and 0.002 respectively) using an unresolved fraction of 0.96 from their Figure 3. The rho, with errors included, becomes 0.71.

\begin{figure}
    \centering
    \includegraphics[width=0.5\textwidth]{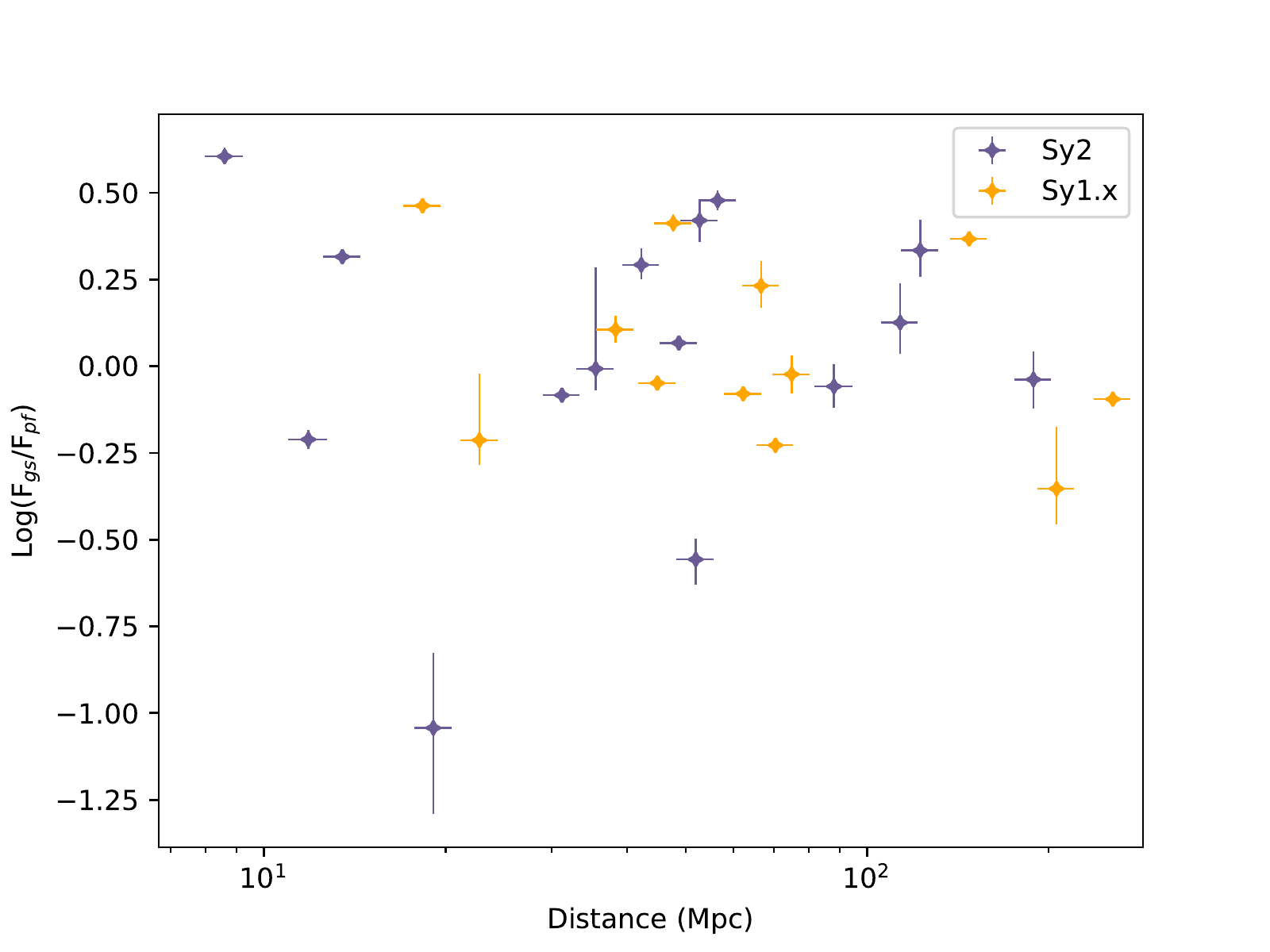}
    \caption{Flux ratio against distance for all objects, including NGC\,1052.}
    \label{fig:Distrat}
\end{figure}

\begin{figure}
    \centering
    \includegraphics[width=0.5\textwidth]{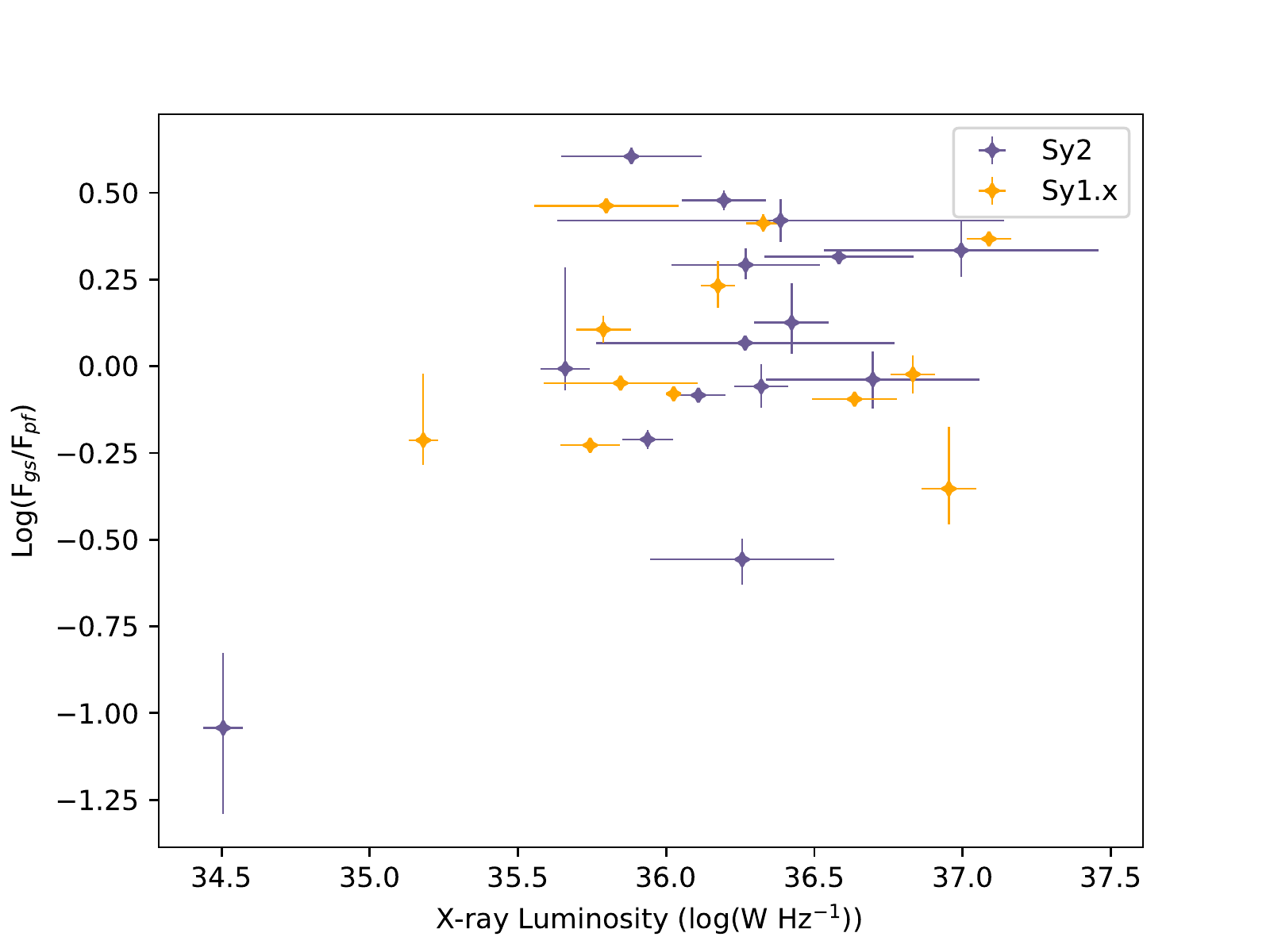}
    \caption{Flux ratio against X-ray luminosity for all objects, including NGC\,1052.}
    \label{fig:XrayLumrat}
\end{figure}

These results are highly suggestive of a positive correlation between the extended source flux fraction and the Eddington ratio. This correlation could, for example, be influenced by the resolution effect. Farther sources would, in theory, be less resolved because these sources would have more of their would-be "extended" emission included in the unresolved emission. Although, we would expect this effect to cause the opposite correlation due to more distant objects generally being more powerful, ergo bright enough to detect with MIDI, and, therefore, likely to be higher Eddington ratio objects. Consequently, we compare the flux ratio to the distance (Figure \ref{fig:Distrat}) and X-ray luminosity (Figure \ref{fig:XrayLumrat}). We find no significant correlation between either of these quantities and the flux ratio. This therefore reinforces the idea that the correlation is truly between flux ratio and Eddington ratio. We discuss this further in the next section.

{}

{}

{}

\section{Discussion}
\label{Discussion}

\subsection{The evolution of dust distribution with Eddington ratio}

In Section \ref{Subsection:eddratres} we presented the results of our comparison between Eddington ratio and the ratio between F$_{\mathrm{gs}}$ and F$_{\mathrm{pf}}$. We found a possible correlation for Sy2s but not for Sy1s, this result can be explained by the orientation of Sy2 allowing us to better distinguish the resolved wind from the compact source. The prediction from \citet{leftley_new_2018} is that at higher Eddington ratios, more of the dust in the system would be located in a more compact disk. This could provide more fuel to the AGN explaining the higher Eddington ratio, and the larger unresolved source fraction.

In this work we find that the unresolved source fraction decreases with Eddington ratio which contradicts this prediction. A z value of $2.4$ for this correlation, when using the MCMC Bayesian statistics directly, means that this correlation is uncertain but possible. This relatively low confidence value is mainly caused by two effects: the large black hole mass errors and the small number of objects. We cannot get better black hole masses due to no near-IR continuum reverberation mapping for Sy2s and the intrinsic scatter in SVD black hole masses. However, we will be able to increase the number of sources thanks to the introduction of MATISSE. With MATISSE a single snapshot analysis of an AGN will provide us with more data, 6 visibility measurements and 3 closure phases, than we have from MIDI on most of our current objects. An improved sensitivity, compared to MIDI, will also allow a greater number of viable targets. This capability should allow a survey of local Sy2s which would vastly improve our statistics.

We repeated our analysis to include a new object, NGC\,1052. This object was studied by \citet{fernandez-ontiveros_compact_2019} and is an obscured LINER. It therefore provides us with an extra observation at low Eddington ratio. We find that our correlation statistics improve to a z of $3.0$ using the Bayesian statistics directly. This demonstrates the possible improvement in our statistics from the inclusion of a small number of new objects. We do not use this object for our final conclusion, however, because it is not clear if the mid-IR emission in this object is from thermally emitting dust. \citet{fernandez-ontiveros_compact_2019} report that the emission is likely synchrotron radiation from a compact jet as a thermally emitting dust structure would have to be unusually compact and the SED follows a broken power-law. The mid-IR emission does show a silicate emission feature, however, and there is an apparent source of nuclear obscuration which implies the presence of dust and dust emission.

In \citet{meisenheimer_resolving_2007}, evidence was given that the unresolved source in Centaurus\,A is also dominated by synchrotron emission while the extended component is dust emission. Similar to NGC\,1052, this object may be unsuitable for our correlation. Ergo, the z value without Centaurus\,A or NGC\,1052 is $2.1$ (p-value of 0.036). This is not significantly different to the result including Centaurus\,A.

To explain the correlation between Eddington ratio and flux ratio we propose a different scenario to that in \citet{leftley_new_2018}. In the radiation driven system of the dusty wind model the higher radiation pressure at higher Eddington ratios would blow more dust out of the disk into the wind. This is further backed by radiation hydrodynamic modelling (see Section \ref{S:rad_mod}).

\subsection{Radiation hydrodynamic simulations of dust distribution}
\label{S:rad_mod}

\citet{williamson_3d_2019} performed radiation hydrodynamics simulations of the production of winds through radiation pressure on a dusty AGN `torus’, using the N-body+Lagrangian hydrodynamics code GIZMO \citep{hopkins_new_2015} in P-SPH mode. These simulations focus on a small spatial scale and a short time-scale, and neglect reprocessed infrared radiation pressure within the torus body, but can provide qualitative insights into the dependence of the distribution of flux on the Eddington ratio. Infrared radiation is important in the structure of the torus but, in most cases, the accretion disk emission is thought to be the main driver of the dusty wind. Therefore, we deem the use of accretion disk emission only as suitable for our purposes. Simulations were performed with Eddington ratios from 0.01 to 0.20. Here we reanalyse the simulations by calculating the infrared emission as a function of radius. We employ a simple model where emission is dominated by dust, dust grains are spherical and emit as black bodies, and do not consider extinction between dust emission and the observer. This simple model is appropriate because we are not generating a high resolution image, but are only interested in the radial dependence of flux.

Temperatures and dust-to-gas ratios were calculated with Cloudy \citep{ferland_2013_2013,ferland_2017_2017} as a function of each particle’s density and temperature, the unextinguished AGN flux received by the particle, and the optical depth from the particle to the AGN. The dust grains have a radius $1\,\mu$m and a density of $3\,$g\,cm$^{-3}$, although the distribution of flux is not strongly dependent on this choice.

The short simulation times (100\,kyr) and lack of explicitly modelled inflow do not permit a steady-state solution to be reached, and so the `torus’ was indeed blown away in these simulations at high Eddington ratios, as suggested in \citet{leftley_new_2018}. However, this is an artificial effect due to the simulations not being reaching dynamical equilibrium. To correct for this effect, we calculate the integrated flux within each radius as a function of $r-r_i$, where $r_i$ is the interior surface of the `torus’. Therefore, if we take the $r_i$ region to be our unresolved component, $r-r_i$ becomes analogous to our extended to unresolved flux ratio. We also perform our analysis at $t=20$\,kyr, before the high Eddington ratio `tori’ can be completely destroyed.

\begin{figure}
    \centering
    \includegraphics[width=0.5\textwidth,trim={0.5cm 0.0cm 0.0cm 0cm},clip]{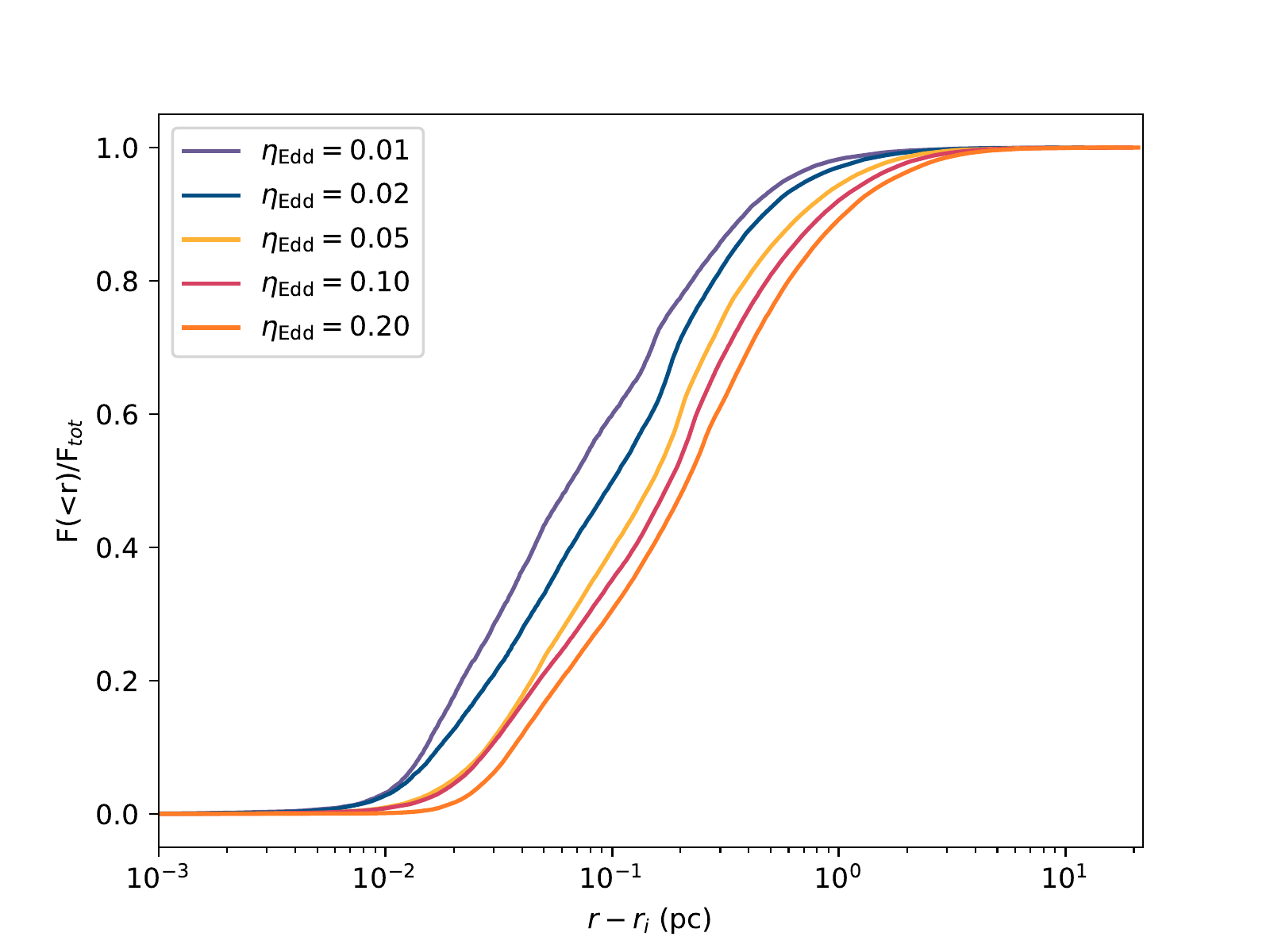}
    \caption{The fraction of flux interior to each radius, as a function of $r-r_i$, where $r$ is the distance from the AGN engine, and $r_i$ is the interior surface of the `torus’, for various Eddington ratios $\eta_\mathrm{Edd}$.}
    \label{fig:David_prediction}
\end{figure}

The results are plotted in Figure \ref{fig:David_prediction}. Here it is clear that the flux becomes {\em less} centrally concentrated with increasing Eddington ratio. This is caused by the outflow rate increasing with greater AGN radiation pressure, producing a stronger and brighter outflow than at lower Eddington ratios. If we extend this qualitative result to the larger scales observed in this work, we can explain the trend in Figure \ref{fig:Eddrat}.

\subsection{MATISSE Candidates}

The coverage and phase information required to reconstruct an image of the dust structure in an AGN from the interferometric data was not available with MIDI. This is expected to change with the second generation instrumentation at the VLTI, and especially with the Multi-AperTure mid-Infrared SpectroScopic Experiment \citep[MATISSE,][]{lopez_overview_2014}. MATISSE will allow the reconstruction of mid-IR images on the same, or smaller, spacial scales as MIDI. MATISSE can access smaller spacial scales using the L and M band, whereas MIDI only had access to the N band.

The unresolved flux of the objects in this work make them viable targets for MATISSE. All the objects with less than 10 observations would benefit from a snapshot observation. However, we especially recommend Fairall\,49, Mrk\,509, and Fairall\,51 due to their tentative partially resolved extended structure. In the case of Fairall\,49 and Mrk\,509, this is the dusty wind component. However, in Fairall\,51 this is the tentative dust "ring".

\section{Summary}
\label{S:Summary}

We reduced and analysed the available MIDI data on 8 AGN. To each we fit a Gaussian and unresolved source model which is the PA independent version of the model used in \citet{leftley_new_2018}. We find that:
\begin{enumerate}
\item All objects are partially resolved, showing visibilities of less than one.

\item Six objects have a unresolved source fraction that can be constrained by the available data. For Mrk\,509 and NGC\,2110 we derived an upper limit for the unresolved source fraction.

\item Two objects have a partially resolved component that can be constrained with the available data; Fairall\,49 and Mrk\,509. In the remaining objects the extended component appears over-resolved so we derive a lower limit for the FWHM.

\item Using the unresolved fraction for all objects, from this paper and the literature, we find a tentative positive correlation between the ratio of extended and unresolved flux and Eddington ratio in Sy2s.

\end{enumerate}
We conclude that this relationship translates to more dust being ejected into the dusty wind at higher Eddington ratios which supports the idea that this wind is a radiation pressure driven outflow. We suggest a survey of Sy2s with VLTI/MATISSE to add more objects to the sample could improve the significance of the tentative correlation.

\acknowledgements
We would like to thank the referee for the thorough
comments and suggestions that helped improved our paper.

JHL, SFH, MV, and DJW acknowledge support from the Horizon 2020 ERC Starting Grant \textit{DUST-IN-THE-WIND} (ERC-2015-StG-677117). MK acknowledges support from JSPS under grant 16H05731. PG acknowledges support from STFC and a UGC/UKIERI Thematic Partnership. DA acknowledges funding through the European Union’s Horizon 2020 and Innovation programme under the Marie Sklodowska-Curie grant agreement no. 793499 (DUSTDEVILS).

Based on European Southern Observatory (ESO) observing programmes 71.B-0062, 078.B-0303, 083.B-0288, 083.B-0452, 084.B-0366, 086.B-0019, 086.B-0242, 087.B-0018, 087.B0401, 091.B-0025, 091.B-0647, 092.B-0718, 092.B-0738, 093.B-0287, 093.B-0647, 094.B-0918, 095.B-0376,0101.B-0411 and 290.B-5113 as well as the Spitzer programmes 86, 526, 3269, 3605, 30572, and 50588.

This research made use of Astropy, a community-developed core Python package for Astronomy \citep{astropy_collaboration_astropy:_2013}. 

This research has made use of the NASA/IPAC Extragalactic Database (NED), which is operated by the Jet Propulsion Laboratory, California Institute of Technology, under contract with the National Aeronautics and Space Administration.

\pagebreak

\bibliographystyle{apj.bst}
\bibliography{Zotero.bib}

\appendix
\section{appendix section}

\startlongtable
\begin{deluxetable}{l @{\extracolsep{\fill}} c c c c c c c}
\tablecaption{List of MIDI Observations\label{tab:files}}
\tablewidth{\textwidth}
\tablehead{
\colhead{Date}       &
\colhead{Time}         &
\colhead{Notes}         &
\colhead{No Track}         &
\colhead{Programme}         &
\colhead{\#Files}     &    
\colhead{PA (deg)}          &
\colhead{BL (m)}}




\startdata
\textbf{Fairall\,51}\\ \hline
2010-10-19&00:04:05&&n&086.B-0019(B)&2&63&73.3\\
2010-10-20&00:09:14&Poor tracking$^{ab}$, included&n&086.B-0019(C)&1&119&82.4\\
2013-06-21&01:50:34&Off centre target&n&091.B-0025(K)&3& 20.0 & 130.2\\
2013-06-21&02:44:46&Fringe tracking failed$^a$&n&091.B-0025(K)&2& 32.0 & 130.0\\
2013-06-23&01:54:23&Fringe tracking failed$^{ab}$&n&091.B-0025(H)&1& 9 & 44.6\\
2013-08-19&04:35:40&Fringe tracking failed$^{ab}$&n&091.B-0647(A)&1& 146 & 62.3\\
2013-08-19&04:51:42&&y&091.B-0647(A)&1& 150 & 62.2\\
2014-05-16&03:00:53&&y&093.B-0287(A)&2& 4 & 130.2\\
2014-05-16&04:10:22&&y&093.B-0287(A)&2& 32 & 86.1\\
2014-05-16&04:56:02&&y&093.B-0287(A)&2& 43 & 87.3\\
2014-05-16&06:51:08&&y&093.B-0287(A)&3& 70 & 89.4\\
2014-05-16&09:14:33&&y&093.B-0287(A)&2& 102 & 86.4\\
2014-05-16&10:02:03&&y&093.B-0287(A)&3& 113 & 83.9\\
2014-08-07&00:58:34&Poor fringe tracking$^{ac}$&n&093.B-0287(E)&1& 48 & 128.5\\
2014-08-07&01:13:28&Insufficient number of central fringes&y&093.B-0287(E)&2& 51 & 128.0\\
2014-08-07&02:38:41&&y&093.B-0287(E)&3& 69 & 123.0\\
2014-08-08&00:24:57&&y&093.B-0287(E)&3 & 14 & 49.8\\
2014-08-08&05:41:38&&y&093.B-0287(E)&2& 130 & 79.6\\
2014-08-08&23:52:17&Low S/N, short exposure time&n&093.B-0287(E)&2& 9 & 50.1\\
2014-08-09&00:05:43&&y&093.B-0647(A)&2& 11 & 50.0\\
2014-08-09&00:54:40&&y&093.B-0647(A)&2& 19 & 49.3\\
2014-08-09&01:24:55&&y&093.B-0647(A)&1& 24 & 48.7\\
2014-08-09&05:33:02&&y&093.B-0647(A)&2& 62 & 36.5\\
\hline
\textbf{Fairall\,49}\\ \hline
2009-08-03&05:30:13&Very high S/N&n&083.B-0288(A)&2& 57 & 38.9\\
2009-08-04&05:29:50&Very high S/N&n&083.B-0288(C)&2& 67 & 69.9\\
2013-06-21&03:18:09&No fringes$^a$, seeing of 1.89&n&091.B-0025(K)&1& 40 & 129.9\\
\hline
\textbf{Fairall\,9}\\ \hline
2013-12-16&00:38:44&No fringes, seeing of 1&y&092.B-0738(D)&1&42&47.6\\
2013-12-17&00:54:48&OPD>100$\mu$m&y&092.B-0738(C)&1&88&89.8\\
2013-12-17&01:33:03&OPD>100$\mu$m&y&092.B-0738(C)&1&87&98.0\\
2013-12-17&03:06:08&OPD>100$\mu$m&y&092.B-0738(C)&1&81&119.5\\
2014-08-08&04:42:40&&n&093.B-0287(E)&2&84&23.7\\
2014-08-08&05:05:15&&y&093.B-0287(E)&2&84&29.5\\
2014-08-09&10:36:28&&y&093.B-0647(A)&2&44&45.4\\
\hline
\textbf{MCG-06-30-15}\\ \hline
2010-03-01&03:49:58&Poor fringe tracking$^a$&n&084.B-0366(A)&1& 175 & 56.6\\
2010-03-01&04:35:22&&n&084.B-0366(A)&4& 2 & 56.6\\
2010-03-02&06:46:10&Insufficient number of central fringes$^{ad}$&n&084.B-0366(C)&2& 27 & 102.3\\
2010-03-03&03:48:11&&n&084.B-0366(B)&1& 79 & 40.7\\
2014-05-15&04:46:12&Failure in fringe tracking$^{a}$, partial tracking only&n&093.B-0287(A)&1& 99 & 80.3\\
2014-05-15&05:08:16&&y&093.B-0287(A)&2& 102 & 76.9\\
2014-05-15&06:03:23&&y&093.B-0287(A)&1& 113 & 66.9\\
2014-05-16&00:57:54&&y&093.B-0287(A)&2& 45 & 127.1\\
2014-05-16&02:06:01&&y&093.B-0287(A)&2& 56 & 130.2\\
2014-05-16&05:53:57&&y&093.B-0287(A)&4& 111 & 68.0\\
\hline
\textbf{Mrk\,509}\\ \hline
2014-08-07&01:40:17&Insufficient number of central fringes&y&093.B-0287(E)&2& 42 & 101.3\\
2014-08-07&03:33:21&Insufficient number of central fringes&y&093.B-0287(E)&2& 57 & 124.3\\
2014-08-07&04:29:20&Insufficient number of central fringes&y&093.B-0287(E)&2& 61 & 129.6\\
2014-08-08&01:28:26&&y&093.B-0287(E)&2& 4 & 51.6\\
2014-08-08&01:39:32&&n&093.B-0287(E)&1& 6 & 51.7\\
2014-08-08&03:55:01&Suppressed 12 micron emission$^{ac}$&n&093.B-0287(E)&1& 80 & 88.6\\
2014-08-08&04:12:05&&y&093.B-0287(E)&2& 80 & 89.2\\
\hline
\textbf{NGC\,2110}\\ \hline
2013-12-14&02:36:09&&y&092.B-0738(B)&2& 107 & 55.0\\
2013-12-14&03:28:23&Insufficient number of central fringes&y&092.B-0738(B)&1& 107 & 60.1\\
2013-12-14&04:07:39&&y&092.B-0738(B)&2& 108 & 62.0\\
2013-12-14&04:56:54&&y&092.B-0738(B)&2& 109 & 62.2\\
2013-12-14&05:36:26&Insufficient number of central fringes&y&092.B-0738(B)&1& 111 & 60.6\\
2013-12-15&02:27:37&Insufficient number of central fringes/high background$^{b}$&y&092.B-0738(A)&2& 7 & 50.4\\
2013-12-15&04:54:22&Insufficient number of central fringes/high background$^{b}$&y&092.B-0738(A)&2& 26 & 54.4\\
2013-12-15&05:36:05&High background$^{b}$&y&092.B-0738(A)&1& 30 & 55.6\\
2013-12-16&02:14:24&High background$^{b}$&y&092.B-0738(D)&1& 20 & 38.6\\
2013-12-16&02:44:45&Insufficient number of central fringes/high background$^{b}$&y&092.B-0738(D)&4& 26 & 39.9\\
2014-11-05&09:01:36&&y&	094.B-0918(B)&3& 115 & 56.3\\
\hline
\textbf{NGC\,7213}\\ \hline
2013-08-18&04:45:11&Insufficient fringes, too faint to track&n&091.B-0647(A)&1& 106 & 61.3\\
2013-08-18&05:28:56&&y&091.B-0647(A)&1& 114 & 62.3\\
2013-08-18&06:30:12&No fringes&y&091.B-0647(A)&2& 125 & 62.2\\
2013-08-18&08:01:22&&y&091.B-0647(A)&2& 144 & 60.0\\
2013-08-18&09:03:47&&y&091.B-0647(A)&2& 159 & 58.2\\
2013-08-19&05:50:13&&y&091.B-0647(A)&2& 119 & 62.5\\
2013-10-20&02:37:37&Low S/N , spike in seeing from clouds&y&092.B-0718(D)&1& 47 & 88.5\\
2013-12-15&01:07:59&Insufficient number of central fringes/high background$^{b}$&y&092.B-0738(A)&1& 52 & 39.4\\
2014-08-09&02:38:57&Insufficient number of central fringes&y&093.B-0647(A)&2& 45 & 81.6\\
2014-08-09&06:34:44&&y&093.B-0647(A)&3& 35 & 51.3\\
2014-08-09&08:10:11&No fringes&y&093.B-0647(A)&1& 87 & 106.2\\
2014-08-09&09:47:50&&y&093.B-0647(A)&3& 53 & 37.8\\
2014-11-03&02:08:41&Insufficient number of central fringes&y&094.B-0918(C)&2& 83 & 111.2\\
\hline
\textbf{NGC\,7674}\\ \hline
2013-08-19&07:23:06&Insufficient number of central fringes&y&091.B-0647(A)&2& 106 & 55.9\\
2013-08-19&08:14:06&Insufficient number of central fringes&y&091.B-0647(A)&3& 107 & 48.8\\
2013-08-19&09:46:12&&y&091.B-0647(A)&4& 112 & 31.0\\
2013-10-20&00:13:28&Insufficient fringes$^{ab}$, very faint&n&092.B-0718(D)&3& 23 & 74.6\\
2013-10-20&00:31:26&Insufficient number of central fringes&y&092.B-0718(D)&2& 26 & 76.9\\
2013-10-20&01:19:37&Insufficient number of central fringes&y&092.B-0718(D)&3& 32 & 83.4\\
2014-08-09&08:57:45&&y&093.B-0647(A)&2& 60 & 128.0\\
\enddata

\tablecomments{a) Due to short coherence time (<4ms), b) high atmospheric dust, c) clouds, d) high wind (pointing limited)}
\end{deluxetable}

\begin{figure*}
    \centering
    \includegraphics[width=\textwidth]{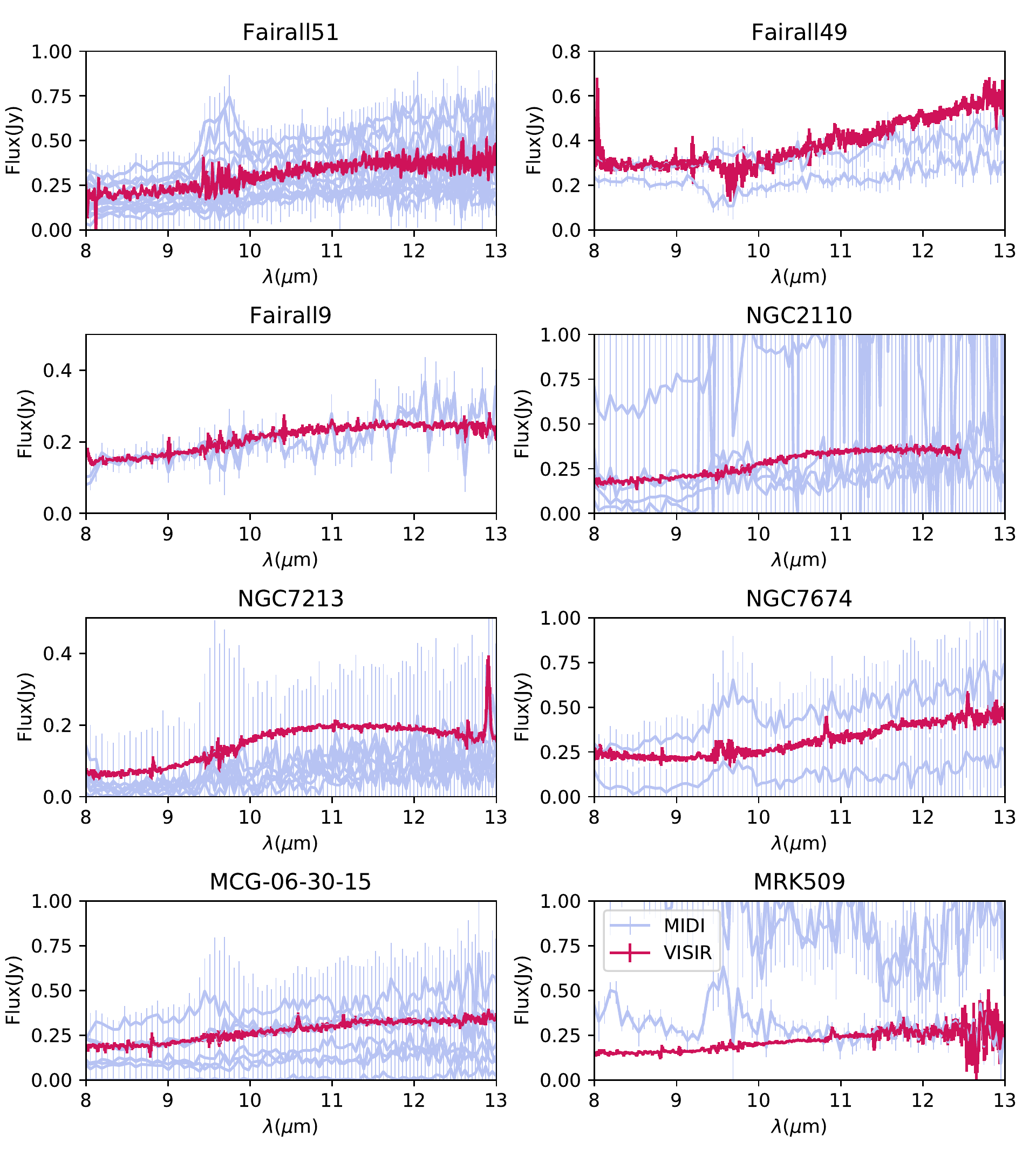}
    \caption{The comparison between the spectrophotometry from VISIR (in red) and MIDI (in light blue). Note that the VISIR spectra of NGC\,7213 and NGC\,2110 have been re-scaled as described in Section\,\ref{Section:VISIR}.}
    \label{fig:Flux comparison}
\end{figure*}

\pagebreak

\begin{table*}[b]
\caption{Summary table of AGN featured in this work}
    \centering
    \addtolength{\tabcolsep}{-2pt}
    \begin{tabular}{l|ccccccccccc}
        \hline
         Object&Type&Ref$^\mathrm{a}$&D$^\mathrm{b}$&L$_{12\,\mu \mathrm{m}}^\mathrm{c}$&L$_{2-10\,\mathrm{Kev}}^\mathrm{d}$&M${_\mathrm{BH}}$&Ref$^\mathrm{e}$ &$\mathrm{N_H}^\mathrm{f}$&F$_\mathrm{gs}^\mathrm{g}$&F$_\mathrm{pf}^\mathrm{h}$&Ref$^\mathrm{i}$ \\
         &&&(Mpc)&(log(W\,Hz$^{-1}$))&(log(W\,Hz$^{-1}$))&(log(M$_\odot$))&& (log(cm$^{-1}$))&(Jy)&(Jy)\\ \hline \hline

NGC\,424&2&$\alpha$&$48.7\pm3.4$&$36.66^{+0.019}_{-0.019}$&$36.3\pm0.5$&$6.97^{+0.25}_{0.25}$&1&$\geqslant25$&$0.35^{0.01}_{0.01}$&$0.3^{0.01}_{0.01}$&Lo\\
NGC\,1068&2&$\beta$&$13.5\pm1.0$&$36.94^{+0.017}_{-0.018}$&$36.6\pm0.25$&$6.93^{+0.36}_{0.36}$&2&$\geqslant25$&$10.81^{0.19}_{0.19}$&$5.23^{0.19}_{0.19}$&Lo\\
NGC\,3783&1.5&$\alpha$&$47.7\pm3.4$&$36.67^{+0.024}_{-0.025}$&$36.33\pm0.059$&$7.37^{+0.077}_{0.076}$&3&$\geqslant20.9$&$0.49^{0.01}_{0.01}$&$0.19^{0.01}_{0.01}$&Lo\\
Circinus&2&$\alpha$&$8.6\pm0.6$&$36.4^{+0.024}_{-0.025}$&$35.9\pm0.24$&$6.23^{+0.09}_{0.09}$&4&$24.7\pm0.17$&$8.08^{0.11}_{0.11}$&$2.01^{0.11}_{0.11}$&Lo\\
NGC\,5506&2&$\alpha$&$31.1\pm2.2$&$36.48^{+0.012}_{-0.012}$&$36.11\pm0.092$&$7.55^{+0.25}_{0.25}$&5&$22.4\pm0.1$&$0.47^{0.01}_{0.01}$&$0.57^{0.01}_{0.01}$&Lo\\
MCG-5-23-16&2&$\alpha$&$42.1\pm3.0$&$36.5^{+0.043}_{-0.047}$&$36.3\pm0.25$&$7.53^{+0.25}_{0.25}$&5&$22.1\pm0.07$&$0.39^{0.01}_{0.01}$&$0.2^{0.02}_{0.02}$&Lo\\
NGC\,4507&2&$\alpha$&$56.5\pm4.0$&$36.79^{+0.028}_{-0.03}$&$36.2\pm0.14$&$7^{+0.25}_{0.25}$&6&$23.7\pm0.19$&$0.48^{0.01}_{0.01}$&$0.16^{0.01}_{0.01}$&Lo\\
1Zw1&1&$\beta$&$255.1\pm17.9$&$37.91^{+0.34}_{-0.301}$&$36.6\pm0.14$&$7.44^{+0.12}_{0.12}$&7&$21\pm0.45$&$0.11^{0.13}_{0.03}$&$0.31^{0.03}_{0.13}$&Bu\\
NGC\,1365&1.8&$\alpha$&$22.7\pm1.6$&$35.66^{+0.014}_{-0.015}$&$35.3\pm0.17$&$7.1^{+0.25}_{0.25}$&5&$23.2\pm0.45$&-\,-&$0.3^{0.01}_{0.01}$&Bu\\
IRAS\,05189-2524&2&$\alpha$&$189\pm13.2$&$37.69^{+0.074}_{-0.089}$&$36.7\pm0.36$&$7.86^{+0.5}_{0.5}$&8&$22.8\pm0.02$&$0.22^{0.03}_{0.03}$&$0.24^{0.03}_{0.03}$&Bu\\
H\,0557-385&1.2&$\alpha$&$151\pm10.9$&$37.45^{+0.094}_{-0.09}$&$36.8\pm0.19$&-\,-&-\,-&$22.8\pm0.78$&$0.13^{0.03}_{0.02}$&$0.28^{0.02}_{0.03}$&Bu\\
IRAS\,09149-6206&1&$\alpha$&$256\pm17.9$&$37.96^{+0.066}_{-0.078}$&$37.0\pm0.25$&-\,-&-\,-&$21.7\pm0.45$&$0.13^{0.02}_{0.02}$&$0.34^{0.02}_{0.02}$&Bu\\
Mrk\,1239&1n&$\alpha$&$93\pm6.6$&$37.17^{+0.037}_{-0.04}$&$36.3\pm0.25$&$8.48^{+0.25}_{0.25}$&5&$23.5\pm0.45$&-\,-&$0.57^{0.05}_{0.05}$&Bu\\
NGC\,3281&2&$\alpha$&$52\pm3.7$&$36.41^{+0.06}_{-0.07}$&$36.3\pm0.3$&$7.26^{+0.25}_{0.25}$&9&$24.3\pm0.45$&$0.07^{0.01}_{0.01}$&$0.25^{0.01}_{0.01}$&Bu\\
NGC\,4151&1.5&$\alpha$&$18.3\pm1.3$&$36.01^{+0.015}_{-0.016}$&$35.8\pm0.24$&$7.56^{+0.051}_{0.047}$&10&$22.7\pm0.19$&$0.87^{0.01}_{0.01}$&$0.3^{0.01}_{0.01}$&Bu\\
3C273&1&$\alpha$&$705\pm49.4$&$38.68^{+0.051}_{-0.058}$&$38.8\pm0.1$&$8.84^{+0.077}_{0.113}$&10&$\geqslant19.7$&-\,-&$0.32^{0.04}_{0.04}$&Bu\\
NGC\,4593&1&$\alpha$&$45\pm3.2$&$36.18^{+0.1}_{-0.13}$&$35.8\pm0.26$&$6.88^{+0.84}_{0.104}$&11&$20.4\pm0.45$&$0.12^{0.03}_{0.03}$&$0.13^{0.01}_{0.01}$&Bu\\
Centaurus\,A&2&$\gamma$&$12\pm0.9$&$35.79^{+0.026}_{-0.028}$&$35.94\pm0.086$&$7.74^{+0.073}_{0.073}$&12&$23.1\pm0.45$&$0.56^{0.03}_{0.03}$&$0.91^{0.03}_{0.03}$&Bu\\
IRAS\,13349+2438&1n&$\alpha$&$480\pm33.6$&$38.54^{+0.017}_{-0.018}$&$36.89\pm0.075$&-\,-&-\,-&$21.6\pm0.45$&-\,-&$0.5^{0.02}_{0.02}$&Bu\\
IC\,4329A&1.2&$\alpha$&$75\pm5.3$&$37.28^{+0.051}_{-0.058}$&$36.83\pm0.076$&$7^{+0.463}_{0.463}$&10&$21.5\pm0.42$&$0.55^{0.05}_{0.05}$&$0.58^{0.05}_{0.05}$&Bu\\
NGC\,5995&2&$\delta$&$113\pm8.0$&$37.1^{+0.084}_{-0.11}$&$36.4\pm0.13$&$7.11^{+0.4}_{0.4}$&13&$22\pm0.45$&$0.18^{0.03}_{0.02}$&$0.15^{0.02}_{0.03}$&Bu\\
NGC\,7469&1.5&$\alpha$&$67\pm4.7$&$36.92^{+0.061}_{-0.071}$&$36.17\pm0.059$&$6.96^{+0.048}_{0.05}$&10&$\geqslant24.4$&$0.39^{0.03}_{0.03}$&$0.23^{0.03}_{0.03}$&Bu\\
Mrk\,509&1.5&$\alpha$&$148\pm10.4$&$37.24^{+0.2}_{-0.61}$&$37.09\pm0.076$&$8.05^{+0.035}_{0.035}$&10&$\geqslant20.7$&$0.28^{0.058}_{0.076}$&$0.14^{0.07}_{0.09}$&Th\\
Fairall\,49&2&$\alpha$&$88\pm6.4$&$37.07^{+0.043}_{-0.048}$&$36.32\pm0.092$&$7.38^{+0.25}_{0.25}$&14&$22.2\pm0.45$&$0.26^{0.02}_{0.02}$&$0.26^{0.02}_{0.02}$&Th\\
Fairall\,51&1&$\epsilon$&$62\pm4.4$&$36.64^{+0.013}_{-0.014}$&$36.03\pm0.025$&$6.84^{+0.5}_{0.5}$&13&$22.4\pm0.12$&$0.18^{0.004}_{0.004}$&$0.21^{0.004}_{0.004}$&Th\\
MCG-06-30-15&1&$\epsilon$&$38.3\pm2.7$&$36.16^{+0.036}_{-0.04}$&$35.79\pm0.092$&$5.93^{+0.25}_{0.25}$&9&$20.2\pm0.45$&$0.19^{0.01}_{0.01}$&$0.14^{0.009}_{0.009}$&Th\\
NGC\,2110&2&$\alpha$&$35.5\pm2.5$&$36.13^{+0.15}_{-0.23}$&$35.66\pm0.084$&$8.4^{+0.25}_{0.25}$&5&$22.5\pm0.06$&$0.2^{0.05}_{0.05}$&$0.16^{0.05}_{0.05}$&Th\\
NGC\,7213&1.5&$\zeta$&$22.8\pm1.6$&$35.46^{+0.061}_{-0.061}$&$35.18\pm0.05$&$7.3^{+0.25}_{0.25}$&5&$20.3\pm0.45$&$0.07^{0.01}_{0.007}$&$0.12^{0.008}_{0.01}$&Th\\
NGC\,7674&2&$\alpha$&$122\pm8.6$&$37.27^{+0.069}_{-0.082}$&$37.0\pm0.46$&$7.09^{+0.25}_{0.25}$&5&$\geqslant24.4$&$0.3^{0.02}_{0.02}$&$0.13^{0.02}_{0.02}$&Th\\
ESO\,323-G77&1.2&$\alpha$&$70\pm5.0$&$36.74^{+0.13}_{-0.19}$&$35.7\pm0.1$&$7.39^{+0.5}_{0.5}$&13&$23.6\pm0.45$&$0.14^{0.036}_{0.035}$&$0.23^{0.06}_{0.06}$&Le\\
Fairall\,9&1.2&$\alpha$&$206\pm14.4$&$37.55^{+0.06}_{-0.07}$&$36.95\pm0.092$&$8.3^{+0.078}_{0.116}$&10&$\geqslant20.5$&$0.08^{0.01}_{0.01}$&$0.19^{0.01}_{0.01}$&Th\\
IC\,3639&2&$\alpha$&$52.7\pm3.7$&$36.42^{+0.16}_{-0.26}$&$36.4\pm0.75$&$5.96^{+0.25}_{0.25}$&9&$\geqslant25$&$0.24^{0.02}_{0.02}$&$0.09^{0.009}_{0.009}$&F8\\
NGC\,1052&3h&$\alpha$&$19.1\pm1.4$&$35.13^{+0.48}_{-0.11}$&$34.51\pm0.067$&$7.98^{+0.37}_{0.37}$&15&$23\pm0.45$&$0.005^{0.01}_{0.001}$&$0.118^{0.01}_{0.01}$&F9\\
    \end{tabular}
    \addtolength{\tabcolsep}{2pt}
    \tablecomments{The Luminosities and black hole masses used in this work. a) Reference for given AGN activity type: $\alpha$) \citet{veron-cetty_catalogue_2006}, $\beta$) \citet{osterbrock_spectroscopic_1993}, $\gamma$) \citet{krimm_swift/bat_2013}, $\delta$) \citet{panessa_unabsorbed_2002}, $\epsilon$) \citet{bennert_size_2006}, $\zeta$) \citet{veron-cetty_vizier_1998}; b) Hubble Distances (CMB) from NED; c) Total 12\,$\mu$m Luminosity values derived fluxes provided by \citet{asmus_subarcsecond_2014}; d) Absorption corrected X-ray Luminosity derived from \citet{asmus_subarcsecond_2015}; e) Reference for black hole mass: 1) \citet{gu_emission-line_2006}, 2) \citet{lodato_non-keplerian_2003}, 3) \citet{onken_mass_2002}, 4) \citet{greenhill_warped_2003}, 5) \citet{makarov_hyperleda._2014}, 6) \citet{cid_fernandes_star_2004}, 7) \citet{vestergaard_determining_2006}, 8) \citet{wang_unified_2007}, 9) \citet{garcia-rissmann_atlas_2005}, 10) \citet{peterson_central_2004}, 11) \citet{denney_mass_2006}, 12) \citet{cappellari_mass_2009}, 13) \citet{wang_unified_2007}, 14) \citet{koss_bat_2017} ,15) \citet{ho_search_2009}; f) N$_\mathrm{H}$ values from \citet{asmus_subarcsecond_2015}; g) extended component flux; h) unresolved source flux; i) Reference for eextended and unresolved fluxes: Lo) \citet{lopez-gonzaga_mid-infrared_2016}, Bu \citet{burtscher_diversity_2013}, Th) This work, Le) \citet{leftley_new_2018}, F8) \citet{fernandez-ontiveros_embedded_2018}, F9) \citet{fernandez-ontiveros_compact_2019}}
    \label{tab:Summary}
\end{table*}

\clearpage

\startlongtable
\begin{deluxetable*}{l @{\extracolsep{\fill}} c c c c}
\tablecaption{List of ISAAC Observations\label{tab:ISAAC}}
\tablewidth{\textwidth}
\tablehead{
\colhead{Object}       &
\colhead{Date}       &
\colhead{Time}         &
\colhead{Programme} &
\colhead{Band} }




\startdata
Fairall\,51& 2013-06-26 & 08:15:25.8863 & 290.B-5113(A) & L   \\
Fairall\,51& 2013-06-26 & 08:16:51.9090 & 290.B-5113(A) & L   \\
Fairall\,51& 2013-06-26 & 08:18:51.3442 & 290.B-5113(A) & M   \\
Fairall\,51& 2013-06-26 & 08:20:18.0000 & 290.B-5113(A) & M   \\
Fairall\,49& 2013-06-26 & 07:49:39.6196 & 290.B-5113(A) & L   \\
Fairall\,49& 2013-06-26 & 07:51:08.0927 & 290.B-5113(A) & L   \\
Fairall\,49& 2013-06-26 & 07:53:09.7430 & 290.B-5113(A) & M   \\
Fairall\,49& 2013-06-26 & 07:54:34.9486 & 290.B-5113(A) & M   \\
Fairall\,9& 2010-10-07 & 02:45:49.1390 & 086.B-0242(B) & M   \\
Fairall\,9& 2010-10-07 & 02:47:16.1162 & 086.B-0242(B) & M   \\
Fairall\,9& 2010-10-15 & 02:18:14.8860 & 086.B-0242(B) & M   \\
Fairall\,9& 2010-10-15 & 02:19:41.1381 & 086.B-0242(B) & M   \\
Fairall\,9& 2010-10-22 & 03:46:13.8033 & 086.B-0242(B) & M   \\
Fairall\,9& 2010-10-22 & 03:47:38.1333 & 086.B-0242(B) & M   \\
Fairall\,9& 2010-10-27 & 01:56:13.4726 & 086.B-0242(B) & M   \\
Fairall\,9& 2010-10-27 & 01:57:38.9651 & 086.B-0242(B) & M   \\
Fairall\,9& 2010-11-10 & 00:14:07.9850 & 086.B-0242(B) & M   \\
Fairall\,9& 2010-11-10 & 00:15:33.5360 & 086.B-0242(B) & M   \\
Fairall\,9& 2010-11-11 & 00:01:09.7147 & 086.B-0242(B) & M   \\
Fairall\,9& 2010-11-11 & 00:02:34.0413 & 086.B-0242(B) & M   \\
Fairall\,9& 2010-11-22 & 01:02:22.1393 & 086.B-0242(B) & M   \\
Fairall\,9& 2010-11-22 & 01:03:48.2911 & 086.B-0242(B) & M   \\
Fairall\,9& 2010-11-26 & 01:13:47.6723 & 086.B-0242(B) & M   \\
Fairall\,9& 2010-11-26 & 01:15:13.2734 & 086.B-0242(B) & M   \\
Fairall\,9& 2010-12-07 & 00:31:34.9685 & 086.B-0242(B) & M   \\
Fairall\,9& 2010-12-07 & 00:33:01.7285 & 086.B-0242(B) & M   \\
Fairall\,9& 2010-12-13 & 00:56:50.4999 & 086.B-0242(B) & M   \\
Fairall\,9& 2010-12-13 & 00:58:14.5939 & 086.B-0242(B) & M   \\
Fairall\,9& 2011-01-09 & 01:16:23.4416 & 086.B-0242(B) & M   \\
Fairall\,9& 2011-01-09 & 01:17:47.7227 & 086.B-0242(B) & M   \\
Fairall\,9& 2011-07-14 & 06:41:46.3601 & 087.B-0018(B) & M   \\
Fairall\,9& 2011-07-14 & 06:43:12.1778 & 087.B-0018(B) & M   \\
Fairall\,9& 2011-07-20 & 09:45:08.6944 & 087.B-0018(B) & M   \\
Fairall\,9& 2011-07-20 & 09:46:33.2340 & 087.B-0018(B) & M   \\
Fairall\,9& 2011-07-21 & 07:52:47.3671 & 087.B-0018(B) & M   \\
Fairall\,9& 2011-07-21 & 07:54:11.5805 & 087.B-0018(B) & M   \\
Fairall\,9& 2011-07-21 & 08:13:06.3934 & 087.B-0018(B) & M   \\
Fairall\,9& 2011-07-21 & 08:14:33.8256 & 087.B-0018(B) & M   \\
Fairall\,9& 2011-08-18 & 09:17:22.2906 & 087.B-0018(B) & M   \\
Fairall\,9& 2011-09-04 & 09:29:24.9498 & 087.B-0018(B) & M   \\
Fairall\,9& 2011-09-04 & 09:30:49.2816 & 087.B-0018(B) & M   \\
Fairall\,9& 2011-09-12 & 09:35:55.6440 & 087.B-0018(B) & M   \\
Fairall\,9& 2011-09-12 & 09:37:22.9466 & 087.B-0018(B) & M   \\
Fairall\,9& 2011-09-15 & 06:54:45.6165 & 087.B-0018(B) & M   \\
Fairall\,9& 2011-09-15 & 06:56:11.3908 & 087.B-0018(B) & M   \\
MCG-6-30-15& 2013-07-04 & 01:51:18.5046 & 290.B-5113(A) & L   \\
MCG-6-30-15& 2013-07-04 & 01:52:44.5524 & 290.B-5113(A) & L   \\
Mrk\,509& 2003-04-21 & 08:11:01.1533 & 71.B-0062(A) & H   \\
Mrk\,509& 2003-04-21 & 08:17:14.8658 & 71.B-0062(A) & H   \\
Mrk\,509& 2003-04-21 & 08:19:52.8216 & 71.B-0062(A) & Ks   \\
Mrk\,509& 2003-04-21 & 08:26:10.1533 & 71.B-0062(A) & Ks   \\
Mrk\,509& 2011-08-02 & 04:02:48.1658 & 087.B-0018(B) & M   \\
Mrk\,509& 2011-08-02 & 04:04:12.1794 & 087.B-0018(B) & M   \\
Mrk\,509& 2011-09-13 & 23:28:42.6615 & 087.B-0018(B) & M   \\
Mrk\,509& 2011-09-13 & 23:30:05.9334 & 087.B-0018(B) & M   \\
Mrk\,509& 2011-09-23 & 23:45:18.2757 & 087.B-0018(B) & M   \\
Mrk\,509& 2011-09-23 & 23:46:42.7027 & 087.B-0018(B) & M   \\
Mrk\,509& 2011-09-23 & 23:50:22.2386 & 087.B-0018(B) & M   \\
Mrk\,509& 2011-09-23 & 23:51:54.4285 & 087.B-0018(B) & M   \\
Mrk\,509& 2011-09-24 & 23:20:24.1184 & 087.B-0018(B) & M   \\
Mrk\,509& 2011-09-24 & 23:21:50.9830 & 087.B-0018(B) & M   \\
Mrk\,509& 2013-07-03 & 05:26:37.4303 & 290.B-5113(A) & L   \\
Mrk\,509& 2013-07-03 & 05:28:02.8710 & 290.B-5113(A) & L   \\
Mrk\,509& 2013-07-03 & 05:30:03.3439 & 290.B-5113(A) & M   \\
Mrk\,509& 2013-07-03 & 05:31:28.8123 & 290.B-5113(A) & M   \\
Mrk\,509& 2013-07-03 & 05:32:47.8805 & 290.B-5113(A) & M   \\
Mrk\,509& 2013-07-03 & 05:34:14.6913 & 290.B-5113(A) & M   \\
NGC\,2110& 2006-11-19 & 08:18:03.5029 & 078.B-0303(B) & Ks   \\
NGC\,2110& 2007-01-03 & 02:51:45.8785 & 078.B-0303(B) & Ks  \\
NGC\,2110& 2007-01-03 & 02:52:35.9237 & 078.B-0303(B) & Ks   \\
NGC\,2110& 2007-01-03 & 02:53:29.9559 & 078.B-0303(B) & Ks   \\
NGC\,2110& 2007-01-03 & 02:54:20.6548 & 078.B-0303(B) & Ks   \\
NGC\,2110& 2007-01-03 & 02:55:09.8860 & 078.B-0303(B) & Ks   \\
NGC\,2110& 2007-01-03 & 02:56:00.0158 & 078.B-0303(B) & Ks \\
NGC\,2110& 2007-01-03 & 03:22:05.0696 & 078.B-0303(B) & Ks \\
NGC\,2110& 2007-01-03 & 03:23:34.4361 & 078.B-0303(B) & Ks\\
NGC\,2110& 2007-01-03 & 03:25:02.9489 & 078.B-0303(B) & Ks\\
\enddata

\end{deluxetable*}

\begin{figure*}
\includegraphics[width=\textwidth]{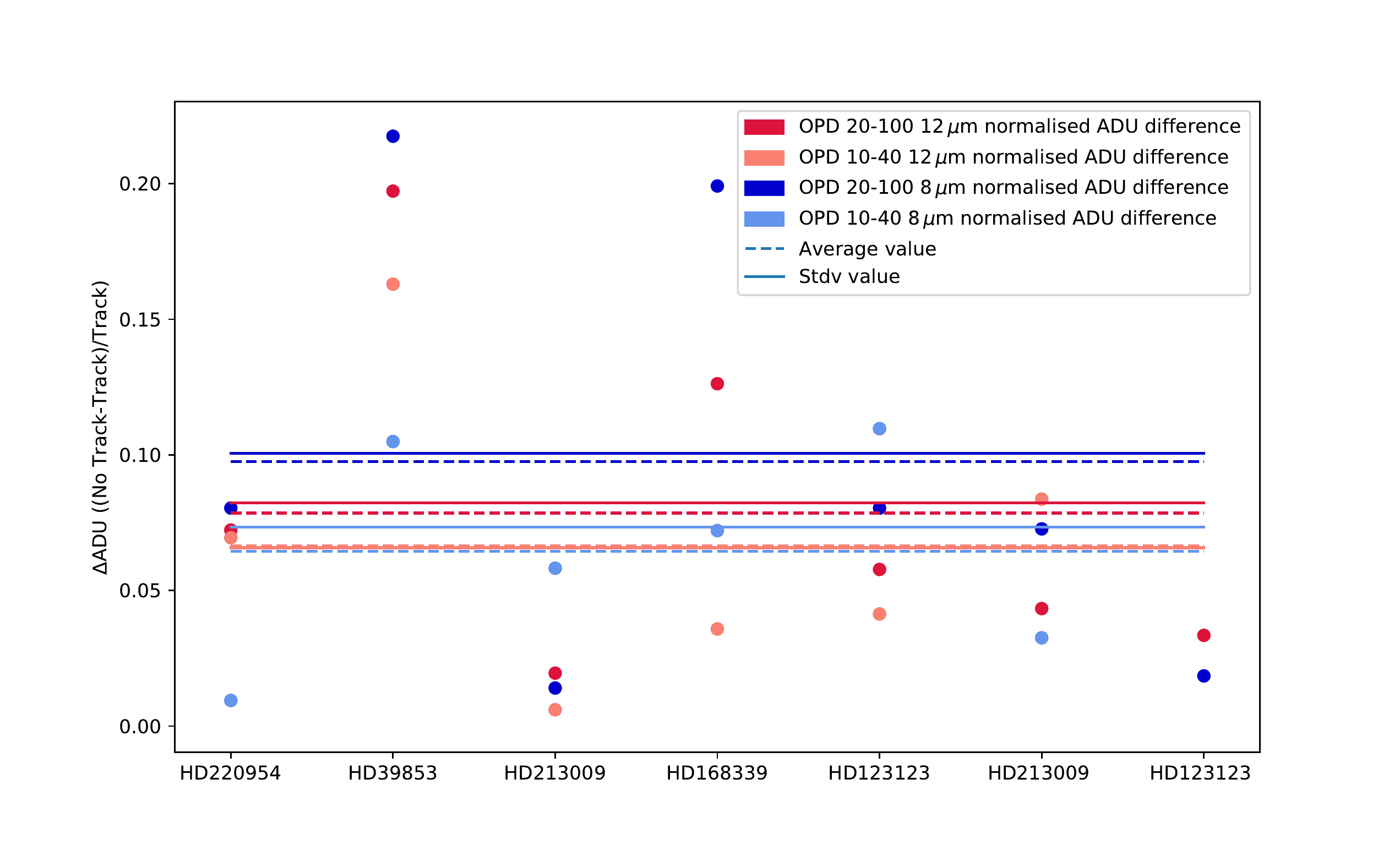}
\caption{Here we show the normalised difference between adjacent tracked and no track observations for 7 calibrators. We plot 2 different OPD ranges for 2 wavelengths and for each pair we plot the average value and standard deviation of the points.}
\label{F: ADU comp multi calib}
\end{figure*}

\begin{figure*}
\includegraphics[width=\textwidth]{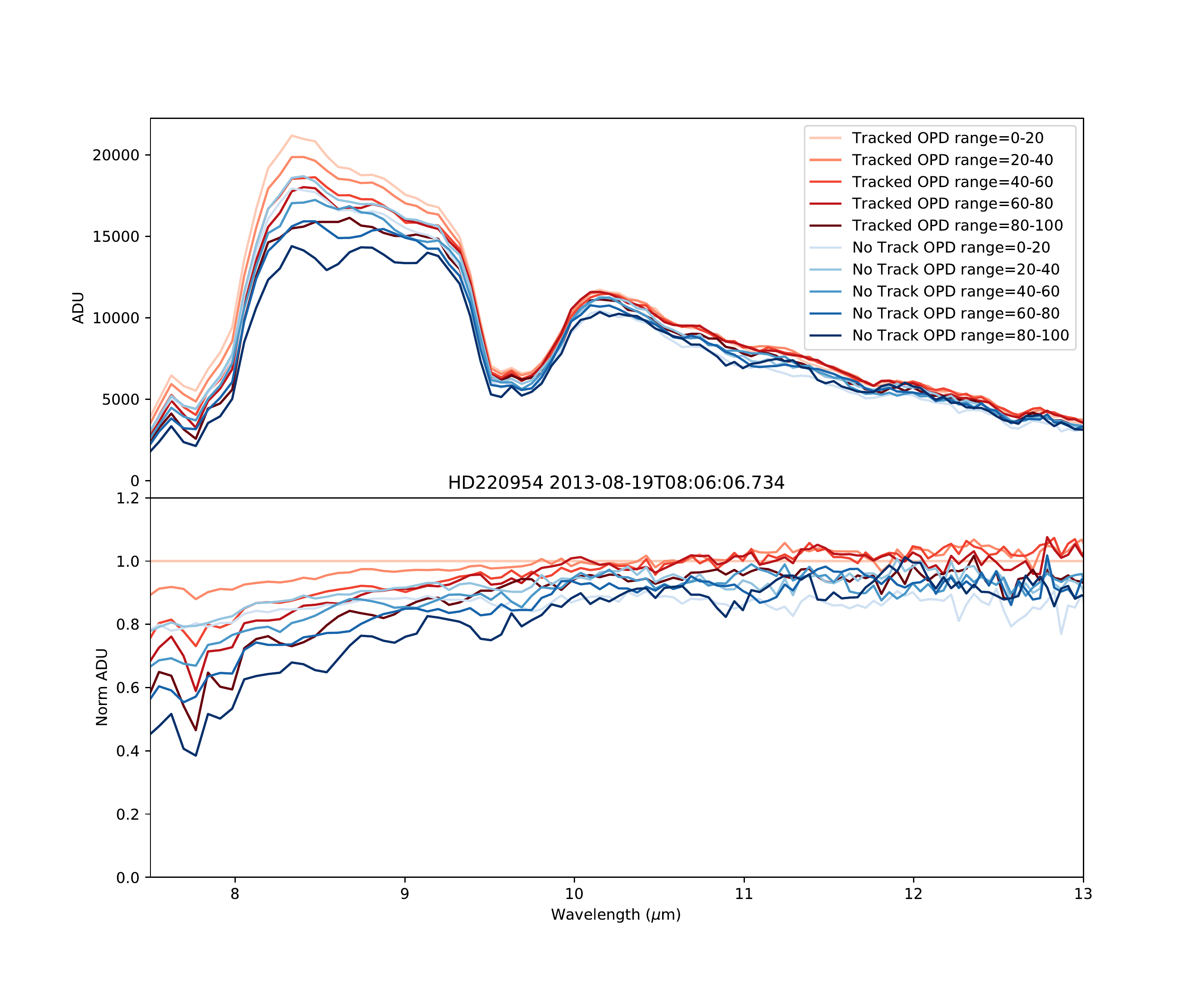}
\caption{Both plots depict the uncalibrated spectrum of HD220954 when reduced with differing OPD ranges. The red lines denote the tracked calibrator and the blue lines denotes the no track. The lower plot is normalised by the Tracked OPD range 0-20 spectrum.}
\label{F: ADU comp}
\end{figure*}

\end{document}